\documentclass[preprint,onecolumn,12pt]{elsarticle}

\usepackage{amssymb, bm}
\usepackage{amsthm, amsmath}

\usepackage{graphicx,color}
\usepackage{caption}
\usepackage{subcaption}
\usepackage{multirow}
\usepackage{mathtools,xparse}

\makeatletter
\newenvironment{smallermatrix}[1][c]
{\null\,\vcenter\bgroup
  \Let@\restore@math@cr\default@tag
  \baselineskip0pt \lineskip0.4pt \lineskiplimit0pt
  \ialign\bgroup\if#1l\else\hfil\fi$\m@th\scriptstyle##$\if#1r\else\hfil\fi&&\thickspace\hfil
  $\m@th\scriptstyle##$\hfil\crcr
}{%
  \crcr\egroup\egroup\,%
}
\makeatother

\NewDocumentCommand{\ts}{O{c} e{^?_}}{
  \begin{smallermatrix}[#1]
  \mathstrut\IfValueT{#2}{#2} \\
  \mathstrut\IfValueT{#3}{#3} \\
  \mathstrut\IfValueT{#4}{#4}
  \end{smallermatrix}%
}

\newcommand{\revise}[1]{{\color{black}{#1}}}

\journal{arXiv}

\begin{document}

\begin{frontmatter}

\title{Proper Orthogonal Descriptors for Multi-element Chemical Systems}

\author[inst1]{Ngoc Cuong Nguyen}

\affiliation[inst1]{organization={Center for Computational Engineering, Department of Aeronautics and Astronautics, Massachusetts Institute of Technology},
            addressline={77 Massachusetts
Avenue}, 
            city={Cambridge},
            state={MA},
            postcode={02139}, 
            country={USA}}

\begin{abstract}
We introduce the  proper orthogonal descriptors for efficient and accurate interatomic potentials of multi-element chemical systems. The potential energy surface of a multi-element system is represented as a many-body expansion of parametrized potentials which are functions of atom positions, atom types, and parameters. The proper orthogonal decomposition is employed to decompose the parametrized potentials {as a linear combination} of orthogonal basis functions. The orthogonal basis functions are used to construct proper orthogonal descriptors based on the elements of atoms, thus leading to multi-element descriptors.  We compose the multi-element proper orthogonal descriptors to develop  linear and quadratic interatomic potentials. We devise an algorithm to efficiently compute the total energy and forces of the interatomic potentials constructed from the proper orthogonal descriptors. The  potentials are demonstrated for indium phosphide and titanium dioxide in comparison with the spectral neighbor analysis potential (SNAP) and  atomic cluster expansion (ACE) potentials. 
\end{abstract}

\begin{highlights}
\item Proper orthogonal descriptors are introduced for multi-element chemical systems. 

\item Two potentials are developed by using linear and quadratic expansions of the proper orthogonal descriptors.

\item The potentials are demonstrated for indium phosphide and titanium dioxide.

\item The quadratic potential is  more accurate than the linear potential at the same cost. 

\end{highlights}


\end{frontmatter}

\section{Introduction}
\label{sec:intro}

Multi-element systems such as alloys, compounds, and composite materials are used in numerous applications and products. The properties of these classes of materials vary with both chemical structures and component compositions. {\em Ab initio} methods for self-consistent calculation of the electronic structures provide the most accurate modeling technique to understanding physical and chemical properties of multi-element systems. As {\em ab initio} methods can be extremely expensive due to their high computational complexity, they are restricted to solving small-scale systems with thousands of atoms at most. Interatomic potentials represent the potential energy surface of  a system as a function of atomic positions and types, leaving out the detailed electronic structures. As most interatomic potentials have a computational complexity that scales linearly with the number of atoms, they  enable molecular dynamics simulations of large-scale systems with millions or even billions of atoms.  

An interatomic potential is essentially a mathematical model that takes the full quantum many-body problem and casts it into computationally tractable expressions for the total energy as the sum of individual atom contributions. However, this simplification can be done in many different ways, giving rise to the numerous different types of interatomic potentials in the literature. The simplest and most efficient potentials are two-body potentials which express the energies and forces solely in terms of pairwise interactions between nearby atoms. Over the years, sophisticated empirical potentials have been developed to treat a wide variety of systems with different degrees of complexity. More often than not, existing empirical potentials  are recalibrated by reparameterizing the existing set of parameters or extending the energy formula by adding terms and then parameterizing the entire potential anew. This results in many different forms and sets of parameters for widely-used empirical potentials such as EAM \cite{Daw1984},  Stillinger-Weber \cite{Stillinger1985}, Tersoff \cite{Tersoff1988}, EDIP \cite{Bazant1997}, REBO \cite{Brenner2002}, ReaxFF \cite{VanDuin2001}.

In recent years, a significant trend has emerged in the form of machine learning (ML) interatomic potentials, where the potential energy surface is described as a composition of local environment descriptors \cite{Bartok2013,Behler2016,DUSSON2022110946}. In order to obtain an accurate and efficient ML potential, the descriptors must meet several requirements.  First and foremost, the descriptors must be invariant with respect to permutation, rotation, translation, and reflection of atoms in the system. Second, the descriptors need to be differentiable with respect to the atomic positions to enable the calculation of analytic gradients for the forces. Third, the descriptors must provide a  detailed structural description of the local atomic environment to produce accurate and transferable potentials. Lastly, since the transformation from atom coordinates onto the descriptors has to be carried out for every atom, the computation of the descriptors has to be fast to produce efficient potentials. Examples of ML potentials include the neural network potential (NNP) \cite{Behler2007,Behler2011,Behler2014}, the Gaussian approximation potential (GAP) \cite{Bartok2010,Fujikake2018,Szlachta2014}, the spectral neighbor analysis potential (SNAP) \cite{Thompson2015,Wood2018}, moment tensor potentials (MTP)  \cite{Novoselov2019,Shapeev2016}, and the atomic cluster expansion (ACE) \cite{Drautz2019,Drautz2020,DUSSON2022110946}.  A recent work \cite{Zuo2020} assesses performance of SNAP, GAP, NNP, MTP potentials on a diverse dataset of bcc and fcc metals, as well as diamond group IV semiconductors.

In a recent work \cite{Nguyen2023}, we introduce a new method to represent atomic neighbourhood environments with the so-called proper orthogonal descriptors (PODs) inspired by the reduced basis method \cite{Barrault2004,Boyaval2009b,Grepl2007a,ARCME,Nguyen2008,Nguyen2023d} for parametrized partial differential equations. The POD potential is expressed as a linear combination of the proper orthogonal descriptors and fitted against density functional theory (DFT) data by using least-squares regression. The POD potential is demonstrated on single-element systems of Li, Mo, Cu, Ni, Si, Ge, Ta elements and found to provide as accurate predictions as both SNAP and ACE potentials. \revise{The POD potential is studied in the recent work \cite{Rohskopf2023} on the influence of model complexity on simulated properties and further extended in \cite{Nguyen2023b} to higher-order interactions. In \cite{Nguyen2023b}, the POD potential is shown to admit the atom density formulation that allows for an efficient implementation of many-body POD descriptors.}

{In this paper, we extend our previous work \cite{Nguyen2023} to develop proper orthogonal descriptors for multi-element chemical systems. The potential energy surface of a multi-element system is represented as a many-body expansion of parametrized potentials which are functions of atom positions, atom types, and parameters. The orthogonal basis functions are formed by applying the proper orthogonal decomposition to the parametrized potentials and transformed into proper orthogonal descriptors according to the chemical elements of atoms, thus leading to multi-element descriptors. Systems with more elements usually have  more complex atomic environments than those with less elements. Hence, increasing the number of elements usually leads to an increase in the number of descriptors in order for the potential to provide accurate predictions for multi-element systems, which in turn results an increase in the computational cost. We emphasize that the multi-element proper orthogonal descriptors only need to scale the size of the model with actual chemical complexity and not with the combinatorial complexity of treating chemical species.}

{To deal with the chemical complexity of multi-element systems, we develop two different POD potentials by composing the descriptors in two different ways. The first POD potential is expressed as a linear combination of the multi-element descriptors. For the second POD potential, we compose the original multi-element descriptors to construct quadratic descriptors that can capture higher-body interactions and provide more accurate predictions for multi-element systems. We develop an algorithm to evaluate the quadratic descriptors and their derivatives without increasing the overall computational cost. The second POD potential has the potential to effectively deal with chemical complexity of multi-element systems because ({\em i}) the cost of evaluating each of its descriptors is independent of the number of elements, ({\em ii}) it is able to generate linear models with sufficiently large number of descriptors, and ({\em iii}) it can retain the computational complexity of the first POD potential. Indeed, when the two POD potentials are demonstrated for indium phosphide and titanium dioxide, the second POD potential is found to be significantly more accurate than the first POD potential for the same computational cost. The performance improvement of the second POD potential over the first POD potential can be attributed to its capability of having considerably more descriptors to capture chemical complexity of multi-element systems.}

{The development of many-body descriptors for multi-element chemical systems have attracted considerable interest \cite{Artrith2017,Gastegger2018,Thompson2015,Willatt2018,Gubaev2019,Darby2022,Kostiuchenko2019,darby2022tensor,Nigam2020,Goscinski2021,Zeni2021}. This interest stems from the wider range of applications and the needs for developing ML potentials that are efficient and accurate even though the atomic systems have a large number of elements. The complexity scaling with the number of elements can restrict the use of  ML potentials in applications with many chemical elements, if the large size and computational cost of  ML models increase rapidly with the number of elements. Various techniques to tackle this scaling problem have
been proposed including element weighting \cite{Artrith2017,Gastegger2018,Thompson2015}, element embedding \cite{Willatt2018,Gubaev2019}, directly reducing the element-sensitive correlation order \cite{Darby2022}, low-rank tensor-train approximations for lattice models \cite{Kostiuchenko2019}, tensor-reduced atomic density representations \cite{darby2022tensor}, and data-driven approach for selecting the most relevant subsets or compressing the original features \cite{Nigam2020,Goscinski2021,Zeni2021}.} 

{In the element weighting approach \cite{Artrith2017,Gastegger2018,Thompson2015}, the number of model coefficients scales linearly with the number of elements and the computational complexity of evaluating the descriptors is independent of the number of elements. While this approach reduces the number of descriptors and the computational cost, it may also reduce the accuracy of the resulting potential compared to the original model. This observation is the main motivation behind the explicit multi-element extension of the element-weighted SNAP potential \cite{Cusentino2020}. In the element embedding approach \cite{Willatt2018,Gubaev2019},  the number of coefficients for the radial basis scales quadratically with the number of elements and the computational complexity of evaluating the descriptors is independent of the number of elements. However, this approach complicates the training process since it can lead to a nonlinear optimization problem for the model coefficients. Different element weighting and element embedding strategies for compressing local atomic neighbourhood descriptors are discussed in \cite{Darby2022}. A tensor-reduced representation of the local atomic environment further reduces the size of the basis by taking element-wise products across the embedded channels rather than a full tensor product \cite{darby2022tensor}.  Although the number of descriptors does not depend on the number of chemical elements, the approach requires the specification of weight matrices for all the embedded channels. In our approach, multi-element descriptors are constructed from the combination of chemical elements without the use of element weighting or embedding.}

The paper is organized as follows. We introduce multi-element proper orthogonal descriptors in Section 2 and develop the two POD potentials in Section 3. In Section 4, we describe an optimization procedure to fit POD potentials. In Section 5, we present numerical results to assess the performance of not only the POD potentials but also the ACE and SNAP potentials. Finally, in Section 6, we make a number of concluding remarks on the results as well as future work.

\section{Multi-Element Proper Orthogonal Descriptors}

\subsection{Parametrized potential energy surface}

We consider a multi-element system of $N$ atoms with $N_{\rm e}$ unique elements. We denote by $\bm r_n$ and $Z_n$ position vector and type of an atom $n$ in the system, respectively. Note that we have $Z_n \in \{1, \ldots, N_{\rm e} \}$, $\bm R = (\bm r_1, \bm r_2, \ldots, \bm r_N) \in \mathbb{R}^{3N}$, and $\bm Z = (Z_1, Z_2, \ldots, Z_N) \in \mathbb{N}^{N}$. The potential energy surface (PES) of the system can be expressed as a many-body expansion of the form
\begin{equation}
\begin{split}
E(\bm R, \bm Z, \bm \eta, \bm \mu) \ = \ & \sum_{i} V^{(1)}(\bm r_i, Z_i, {\bm \eta}, \bm \mu^{(1)} ) + \frac12 \sum_{i,j} V^{(2)}(\bm r_i, \bm r_j, Z_i, Z_j, \bm \eta, \bm \mu^{(2)})  \\
& + \frac16 \sum_{i,j,k} V^{(3)}(\bm r_i, \bm r_j, \bm r_k, Z_i, Z_j, Z_k, \bm \eta, \bm \mu^{(3)}) + \ldots 
\end{split}
\end{equation}
where $V^{(1)}$ is the one-body potential often used for representing external field or energy of isolated elements, and the higher-body potentials $V^{(2)}, V^{(3)}, \ldots$ are symmetric and zero if two or more indices take identical values. The superscript on each potential denotes its body order. Each $q$-body potential $V^{(q)}$  depends on $\bm \mu^{(q)}$  which are sets of parameters to fit the PES. Note that $\bm \mu$ is a collection of all potential parameters $\bm \mu^{(1)}$, $\bm \mu^{(2)}$, $\bm \mu^{(3)}$, etc, and that $\bm \eta$ is a set of hyperparameters. {Each potential $V^{(q)}$ depends on the same hyperparameters.}

In the absence of external fields, the PES should not depend on the absolute position of atoms, but only on the relative positions. This means that the PES can be rewritten as a function of interatomic distances and angles between the bonds as 
\begin{equation}
\label{eq2}
\begin{split}
E(\bm R, \bm Z, \bm \eta, \bm \mu) \ = \ &  \sum_{i} V^{(1)}(Z_i,\bm \eta, \bm \mu^{(1)}) + \frac12 \sum_{i,j} V^{(2)}( r_{ij}, Z_i, Z_j, \bm \eta, \bm \mu^{(2)})  \\
& + \frac16 \sum_{i,j,k} V^{(3)}(r_{ij}, r_{ik}, \theta_{ijk}, Z_i, Z_j, Z_k, \bm \eta, \bm \mu^{(3)}) + \ldots 
\end{split}
\end{equation}
where each  potential is a function of interatomic distances $r_{ij} = |\bm r_i - \bm r_j|$, bond angles $\theta_{ijk}$, potential parameters, and the hyperparameters.  

A wide variety of interatomic potentials such as Lennard-Jones potential, Morse potential, Stillinger-Weber potential, angle potentials, and dihedral potentials have the form (\ref{eq2}). For these empirical potentials, however, both $\bm \eta$ and $\bm \mu$ are fixed to specific values. In other interatomic potentials such as EAM and Tersoff potentials, the many-body interactions are embedded into the terms of a pair potential in which the nature of the interaction is modified by the local environment of the atom via the bond order parameter, coordination number, or electron density. 

Interatomic potentials rely on parameters to learn relationship between atomic environments and interactions.  Since interatomic potentials are approximations by nature, their parameters need to be set to some reference values or fitted against experimental and/or  numerical data by necessity. In simple potentials such as the Lennard-Jones and Morse potential, the parameters can be set to match the equilibrium bond length and bond strength of a dimer molecule or the surface energy of a solid. Many-body potentials often contain many parameters with limited interpretability and need to be optimized by fitting their parameters against a larger set of data. Typically, potential fitting finds optimal parameters, $\bm \mu^*$, to  minimize a certain loss function of the predicted quantities and data.  Since the fitted potential depends on the data set used to fit it, different data sets will yield different optimal parameters and thus different fitted potentials. When fitting the same functional form on $Q$ different data sets, we would obtain $Q$ different optimized potentials, $E(\bm R,\bm Z, \bm \eta, \bm \mu_q^*), 1 \le q \le Q$. These optimized potentials are typically intended to predict properties that are similar to those they are fitted to. Inaccurate predictions may occur when they are used to predict out-of-fitting properties. Consequently, there exist many different sets of optimized parameters for widely-used empirical potentials such as EAM \cite{Daw1984},  Stillinger-Weber \cite{Stillinger1985}, Tersoff \cite{Tersoff1988}, EDIP \cite{Bazant1997}, REBO \cite{Brenner2002}, ReaxFF \cite{VanDuin2001}. 

{Following the philosophy of the reduced basis method \cite{Barrault2004,Boyaval2009b,Grepl2007a,ARCME,Nguyen2008,Nguyen2023d} for parametrized partial differential equations, the parametrized PES can be viewed as a parametric manifold of potential energies} 
\begin{equation}
\mathcal{M} = \{E(\bm R, \bm Z, \bm \eta, \bm \mu) \ | \  \bm \mu \in \Omega^{\bm \mu} \}    
\end{equation}
where $\Omega^{\bm \mu}$ is a parameter domain in which $\bm \mu$ resides. The parametric manifold $\mathcal{M}$ contains potential energy surfaces for all values of $\bm \mu \in \Omega^{\bm \mu}$.  Therefore, the parametric manifold yields a much richer and  more transferable atomic representation than any particular individual PES $E(\bm R, \bm Z, \bm \eta, \bm \mu^*)$. {In what follows, we propose specific forms of the parametrized potentials for  two-body and three-body interactions, and construct corresponding descriptors by using the  proper orthogonal decomposition.}

\subsection{Two-body proper orthogonal descriptors}

We adopt the usual assumption that the direct interaction between two atoms vanishes smoothly when their distance is greater than the outer cutoff distance $r_{\rm max}$. Furthermore, we assume that two atoms can not get closer than the inner cutoff distance $r_{\rm min}$ due to the Pauli repulsion  principle. {The two-body descriptors are constructed upon a set of orthogonal basis functions, $U^{(2)}_i(r, \bm \eta), 1 \le i \le N_{\rm b}^{(2)}$, which are defined in the domain $r \in [r_{\min}, r_{\max}]$ such that
\begin{equation}
\int_{r_{\min}}^{r_{\max}} U^{(2)}_i(r, \bm \eta)  U^{(2)}_j(r, \bm \eta) d r = \delta_{ij}, \qquad i,j = 1,\ldots,  N_{\rm b}^{(2)} .      
\end{equation}
Here $N_{\rm b}^{(2)}$ is the number of basis functions for the two-body interaction. In \cite{DUSSON2022110946}, a general procedure was proposed to construct radial basis functions which are orthogonal with respect to a user-defined inner product. In particular, the orthogonal basis functions are constructed from a set of orthogonal polynomials on $[a, b]$, a coordinate transformation $\xi : [r_{\min}, r_{\max}] \to [a, b]$, a density function $\rho(r)$, and a cut-off function $f_{\rm cut}(r)$. The procedure allows for a wide range of potential optimization of the radial basis by optimizing the coordinate transformation and the density function. Appropriate choices of the coordinate transformation and the density function have been proposed in \cite{DUSSON2022110946}. Our method differs from the approach introduced in \cite{DUSSON2022110946} as follows. Instead of using orthogonal polynomials to construct the radial basis, we will use two families of parametrized radial functions. Our method is also related to the  principal component analysis \cite{Goscinski2021} that generates  optimal radial basis in terms of its ability to linearly compress the information encoded in the variance of the density coefficients. The optimal radial basis can be determined as a contraction of any complete primitive basis, and evaluated efficiently by approximating it with splines. In \cite{Goscinski2021}, the Gaussian type orbital (GTO) basis and the discrete variable representation (DVR) basis are used as complete primitive basis to construct the optimal radial basis.}

{We describe a method to systematically construct orthogonal basis functions from families of parametrized functions. We introduce two families of parametrized radial functions}
\begin{equation}
\label{eq3}
\phi(r, r_{\rm min}, r_{\rm max}, \alpha, \beta)  = \frac{\sin (\alpha \pi x) }{r - r_{\rm min}}, \qquad  \varphi(r, \gamma)  = \frac{1}{r^\gamma} ,    
\end{equation}
where the scaled distance function $x$ is defined below to enrich the two-body manifold   
\begin{equation}
x(r, r_{\rm min}, r_{\rm max}, \beta) = \frac{e^{-\beta(r - r_{\rm min})/(r_{\rm max} - r_{\rm min})} - 1}{e^{-\beta} - 1} .
\end{equation}
We introduce the following function as a convex combination of the two functions in (\ref{eq3})
\begin{equation}
\label{eq7ww}
\psi(r, r_{\rm min}, r_{\rm max}, \alpha, \beta, \gamma, \kappa)  = \kappa \phi(r, r_{\rm min}, r_{\rm max}, \alpha, \beta) + (1- \kappa)  \varphi(r, \gamma) .
\end{equation}
{The function $\phi$ in (\ref{eq3}) is related to the  zeroth spherical Bessel function, while the function $\varphi$ is inspired by the n-m Lennard-Jones potential. Indeed, the 12-6 Lennard-Jones potential $V_{\rm LJ}(r) = \frac{A}{r^{12}} - \frac{B}{r^6}$ can be expressed as $V_{\rm LJ}(r) = A\varphi(r, 12) - B\varphi(r, 6)$. As a result, $V(r) = \sum_{\gamma=1}^{N_\gamma} A_k \varphi(r, \gamma)$ can describe both repulsive and attractive interactions between two atoms for any given large enough $N_\gamma$. Although the parameter $\gamma$ can be real number, we choose a set of consecutive positive integers  $\{1,2,\ldots,  N_\gamma\}$ to compute instances of the parametrized function $\varphi$  by making use of the relation $\varphi(r, \gamma+1) = \varphi(r, \gamma)/r$. Similarly, we choose a set of consecutive integers $\{1, 2, \ldots, N_{\alpha}\}$ for $\alpha$ to generate instances of the parametrized function $\phi$ because it allows us to use the formula $\sin((\alpha+1) \pi x) = \sin(\pi x) U_{\alpha}(\cos (\pi x))$ to compute these instances efficiently, where $U_{\alpha}$ are Chebyshev polynomials of the second kind. We take $N_{\beta}$ values for the parameter $\beta$ such that $\beta_k = (k-1) \beta_{\max}/(P_{\beta}-1)$ for $k = 1, 2, \ldots, N_{\beta}$, where $\beta_{\max} = 4.0$. Hence, the total number of parameter points is $N_{\rm s} = N_\alpha N_\beta + N_\gamma$. Although  $N_\alpha$, $N_\beta$, $N_\gamma$ can be chosen conservatively large, we find that $N_\alpha = 6$, $N_\beta = 3$, $N_\gamma = 12$ (namely, $N_s = 30$) are adequate to enabling the function $\psi$ to characterize a diverse spectrum of two-body interactions within the cut-off interval $(r_{\min}, r_{\max})$.} 

{Our method allows for the addition of other families of parametrized functions (e.g., the Morse potential) to define $\psi$ in (\ref{eq7ww}). More broadly, the method allows for a systematic combination of empirical potentials to define orthogonal descriptors based on the following  parametrized potential}
\begin{equation}
\label{eq6}
W^{(2)}(r_{ij}, \bm \eta, \bm \mu^{(2)})  = f_{\rm c}(r_{ij}, \bm \eta) \psi(r_{ij}, \bm \eta, \bm \mu^{(2)})
\end{equation}
{where $\bm \eta$ includes the inner and outer cut-off radii, $\bm \mu^{(2)}$ consists of the remaining parameters, and $f_c$ is a cut-off function. In this paper, we use the following cut-off function}
\begin{equation}
\label{eq8}
 f_{\rm c}(r_{ij},  r_{\rm min}, r_{\rm max})  =  \exp \left(1 -\frac{1}{\sqrt{\left(1 - \frac{(r-r_{\min})^3}{(r_{\max} - r_{\min})^3} \right)^2 + 10^{-6}}} \right),
\end{equation}
{which ensures the smooth vanishing of the  potential and its derivative beyond $r_{\rm max}$. Other cut-off functions can also be used. Given $N_{\rm s}$ parameter tuples $\bm \mu^{(2)}_\ell, 1 \le \ell \le N_{\rm s}$, we introduce the following set of  snapshots}
\begin{equation}
\label{eq11}
\xi_\ell(r_{ij}, \bm \eta) =  W^{(2)}(r_{ij}, \bm \eta, \bm \mu^{(2)}_\ell),  \quad \ell = 1, \ldots, N_{\rm s} .
\end{equation}
{We employ the proper orthogonal decomposition~\cite{sirovich87:_turbul_dynam_coher_struc_part,willcox02:_balanced_pod,Nguyen2008a} to generate an orthogonal basis set  as follows}
\begin{equation}
\label{eq12}
U^{(2)}_m(r_{ij}, \bm \eta) = \sum_{\ell = 1}^{N_{\rm s}} A_{\ell m}(\bm \eta) \,  \xi_\ell(r_{ij}, \bm \eta), \qquad m = 1, \ldots, N_{\rm b}^{(2)} , 
\end{equation}
{where the matrix $\bm A \in \mathbb{R}^{N_{\rm s} \times N_{\rm s}}$ consists of eigenvectors of the eigenvalue problem 
\begin{equation}
\bm C \bm a = \lambda \bm a \ ,
\label{eq12b}
\end{equation}
with 
$$C_{ij}  = \frac{1}{N_{\rm s}} \int_{r_{\min}}^{r_{\max}} \xi_i(x, \bm \eta) \xi_j(x, \bm \eta) dx, \quad 1 \le i, j \le N_{\rm s} .$$
Note that the  eigenvalues $\lambda_\ell, 1 \le \ell \le N_{\rm s},$ are ordered such that $\lambda_1 \ge \lambda_2 \ge \ldots \ge \lambda_{N_{\rm s}}$, and that the matrix $\bm A$ is pre-computed and stored. Owing to the rapid convergence of the proper orthogonal decomposition,  a small number of orthogonal basis functions is needed to obtain accurate approximation.} 

Finally, the two-body proper orthogonal descriptors at each atom $i$ are computed by summing the orthogonal basis functions over the neighbors of atom $i$ and numerating on the atom types as follows
\begin{equation}
\label{eq12c}
D^{(2)}_{im l(p, q) }(\bm \eta)  = \left\{
\begin{array}{ll}
\displaystyle \sum_{\{j | Z_j = q\}} U^{(2)}_m(r_{ij},  \bm \eta), & \mbox{if } Z_i = p \\
0, & \mbox{if } Z_i \neq p
\end{array} 
\right.   
\end{equation}
 for $1 \le i \le N, 1 \le m \le N_{\rm b}^{(2)}, 1 \le q, p \le N_{\rm e}$. Here $l(p,q)$ is a symmetric index mapping such that  
 \begin{equation}
\label{eq12d}
l(p,q)  = \left\{
\begin{array}{ll}
q + (p-1) N_{\rm e} - p(p-1)/2, & \mbox{if } q \ge p \\
p + (q-1) N_{\rm e} - q(q-1)/2, & \mbox{if } q < p . 
\end{array} 
\right.   
\end{equation}
The number of descriptors per atom is thus $N_{\rm b}^{(2)} N_{\rm e}(N_{\rm e}+1)/2$. {It is important to note that the orthogonal basis functions defined in (\ref{eq12}) do not depend on the atomic numbers $Z_i$ and $Z_j$. Therefore, the cost of evaluating the basis functions and their derivatives with respect to $r_{ij}$ is independent of the number of elements $N_{\rm e}$. } 

\subsection{Three-body proper orthogonal descriptors}

{In order to provide proper orthogonal descriptors for three-body interactions, we introduce a three-body parametrized potential as follows}
\begin{equation}
\label{eq9}
\begin{split}
W^{(3)}(r_{ij}, r_{ik}, \theta_{ijk}, \bm \eta, \bm \mu^{(3)})  =  \psi(r_{ij}, r_{\rm min}, r_{\rm max}, \alpha, \beta, \gamma, \kappa) f_{\rm c}(r_{ij}, r_{\rm min}, r_{\rm max}) \\
\psi(r_{ik}, r_{\rm min}, r_{\rm max}, \alpha, \beta, \gamma, \kappa) f_{\rm c}(r_{ik}, r_{\rm min}, r_{\rm max}) \\
\cos (\sigma \theta_{ijk} + \zeta) 
\end{split}
\end{equation}
where $\sigma$ is the periodic multiplicity, $\zeta$ is the equilibrium angle, $\bm \mu^{(3)} = (\alpha, \beta, \gamma, \kappa, \sigma, \zeta)$.
The potential (\ref{eq9}) provides an angular fingerprint of the atomic environment through the bond angles $\theta_{ijk}$ formed with each pair of neighbors $j$ and $k$.  Compared to the two-body potential (\ref{eq6}), the three-body potential (\ref{eq9}) has two extra parameters $(\sigma, \zeta)$ associated with the angular component. {Ideally, $\zeta$ is the physical equilibrium angle of the molecular system. However, since the function  $\cos(\sigma \theta + \zeta)$ can be approximated well by a linear combination  of Chebyshev polynomials of the first kind, $T_{n-1}(\cos \theta) = \cos((n-1) \theta), n = 1, \ldots, N_{\rm a}$, for large enough $N_{\rm a}$, we take $\zeta$ to be zero and choose $\sigma$ as non-negative integers.}
{Hence, the angular basis functions are given by
\begin{equation}
U^{a}_n(\theta_{ijk}) = \cos ((n-1) \theta_{ijk}), \qquad  n = 1,\ldots, N_{\rm a}, 
\end{equation}
where $N_{\rm a}$ is the number of angular basis functions. The orthogonal basis functions for the parametrized potential (\ref{eq9}) are computed as follows
\begin{equation}
\label{eq14}
U^{(3)}_{mn}(r_{ij}, r_{ik}, \theta_{ijk}, \bm \eta) = U^{(2)}_m(r_{ij}, \bm \eta) U^{(2)}_m(r_{ik}, \bm \eta) U^{a}_n(\theta_{ijk}),
\end{equation}
for $1 \le m \le N_{\rm r}, 1 \le n \le N_{\rm a}$, where  $U^{(2)}_m(r_{ij}, \bm \eta)$ are the two-body orthogonal basis functions. The number of three-body orthogonal basis functions is equal to $N_{\rm b}^{(3)} = N_{\rm r} N_{\rm a}$. 
}

Finally, the three-body proper orthogonal descriptors at each atom $i$ are obtained by summing (\ref{eq14}) over the neighbors $j$ and $k$ of atom $i$ as
\begin{equation}
\label{eq15}
D^{(3)}_{imn \ell(p, q, s)}(\bm \eta)  = \left\{
\begin{array}{ll}
\displaystyle \sum_{\{j | Z_j = q\}} \sum_{\{k | Z_k = s\}} U^{(3)}_{mn}(r_{ij}, r_{ik}, \theta_{ijk}, \bm \eta), & \mbox{if } Z_i = p \\
0, & \mbox{if } Z_i \neq p
\end{array} 
\right.   
\end{equation}
for $1 \le i \le N, 1 \le m \le N_{\rm r}, 1 \le n \le N_{\rm a}, 1 \le q, p, s \le N_{\rm e}$, where
\begin{equation}
\label{eq15b}
\ell(p,q,s)  = \left\{
\begin{array}{ll}
s + (q-1) N_{\rm e} - q(q-1)/2 + (p-1)N_{\rm e}(1+N_{\rm e})/2 , & \mbox{if } s \ge q \\
q + (s-1) N_{\rm e} - s(s-1)/2 + (p-1)N_{\rm e}(1+N_{\rm e})/2, & \mbox{if } s < q . 
\end{array} 
\right.   
\end{equation}
The number of three-body descriptors per atom is thus $N_{\rm b}^{(3)} N_{\rm e}^2(N_{\rm e}+1)/2$. It remains to describe the computational complexity of the proper orthogonal descriptors.

\subsection{Computational complexity of proper orthogonal descriptors}

{We analyze the computational complexity of evaluating the descriptors and their derivatives with respect to $r_{ij}$. It follows from (\ref{eq12}) that 
\begin{equation}
\label{eq120q}
\frac{\partial U^{(2)}_m(r_{ij}, \bm \eta)}{\partial r_{ij} } = \sum_{\ell = 1}^{N_{\rm s}} A_{\ell m}(\bm \eta) \,  \frac{\partial \xi_\ell(r_{ij}, \bm \eta)}{\partial r_{ij}}, \qquad m = 1, \ldots, N_{\rm b}^{(2)} .  
\end{equation}
and from (\ref{eq14}) that
\begin{equation}
\label{eq140}
\begin{split}
\frac{\partial U^{(3)}_{mn}(r_{ij}, r_{ik}, \theta_{ijk}, \bm \eta)}{\partial r_{ij}} & =  \frac{\partial U^{(2)}_m(r_{ij}, \bm \eta)}{\partial r_{ij}}  U^{(2)}_m(r_{ik}, \bm \eta) \cos((n-1) \theta_{ijk}) \\
\frac{\partial U^{(3)}_{mn}(r_{ij}, r_{ik}, \theta_{ijk}, \bm \eta)}{\partial r_{ik}} & =  
U^{(2)}_m(r_{ij}, \bm \eta) \frac{\partial U^{(2)}_m(r_{ik}, \bm \eta)}{\partial r_{ik}} \cos((n-1) \theta_{ijk}), \\
\frac{\partial U^{(3)}_{mn}(r_{ij}, r_{ik}, \theta_{ijk}, \bm \eta)}{\partial \theta_{ijk}} & = -(n-1) U^{(2)}_m(r_{ij}, \bm \eta) U^{(2)}_m(r_{ik}, \bm \eta) \sin((n-1) \theta_{ijk}) .
\end{split}
\end{equation}
We assume that each atom has the same number of neighbors $N_{\rm n}$. The total number of neighbors is thus $N N_{\rm n}$ for all atoms. The operation count for computing the orthogonal basis functions in (\ref{eq12}) and their derivatives in (\ref{eq120q}) is thus $O(N N_{\rm n} N_{\rm s} N_{\rm b}^{(2)})$. The operation count for evaluating the two-body PODs in (\ref{eq12c}) and their derivatives  is  $O(N N_{\rm n} N_{\rm b}^{(2)} )$.  The operation count for computing the three-body orthogonal basis functions in (\ref{eq14}) and their derivatives in (\ref{eq140}) is $O(N N_{\rm n}^2 N_{\rm b}^{(3)})$. Therefore, the computational complexity of the three-body PODs (\ref{eq15}) and their derivatives is also $O(N N_{\rm n}^2 N_{\rm b}^{(3)})$. Since the three-body PODs are more expensive than the two-body PODs, the overall computational complexity is $O(N N_{\rm n}^2 N_{\rm b}^{(3)})$.}

{Of course, systems with more elements often have  more complex atomic environments than those with less elements. Hence, increasing the number of elements may require an increase in the number of descriptors, thereby increasing the computational cost of a potential constructed from these descriptors. To deal with the chemical complexity of multi-element systems, we introduce a quadratic POD potential by composing the multi-element descriptors to construct  quadratic descriptors that can capture higher-body interactions and chemical complexity of multi-element systems.  We develop an algorithm to evaluate the quadratic descriptors and their derivatives without increasing the overall computational cost. The quadratic POD potential has the following features: ({\em i}) the cost of evaluating each of its descriptors is independent of the number of elements, ({\em ii}) it can generate linear models with  large number of descriptors, and ({\em iii}) it can retain the computational complexity of the linear POD potential. The next section describes the construction of the linear and quadratic POD potentials.}

\section{Proper Orthogonal Descriptor Potentials}

\subsection{The linear POD potential}

{We use the proper orthogonal descriptors to define the atomic energies 
\begin{multline}
\label{eq16}
E^{\rm L}_{i}(\bm \eta) = \sum_{p=1}^{N_{\rm e}} c^{(1)}_p D^{(1)}_{ip} + \sum_{m=1}^{N_{\rm b}^{(2)}}  \sum_{l=1}^{N_{\rm e}(N_{\rm e}+1)/2} c^{(2)}_{ml} D^{(2)}_{iml}(\bm \eta)  \, +  \\
\sum_{m=1}^{N_{\rm r}} \sum_{n=1}^{N_{\rm a}}  \sum_{\ell=1}^{N_{\rm e}^2(N_{\rm e}+1)/2} c^{(3)}_{mn\ell} D^{(3)}_{imn\ell}(\bm \eta), \quad 1 \le i \le N, 
\end{multline}
where $D^{(1)}_{ip}, D^{(2)}_{iml}, D^{(3)}_{imn\ell}$ are the  one-body, two-body descriptors, respectively, and $c^{(1)}_p, c^{(2)}_{ml}, c^{(3)}_{mn\ell}$ are their respective expansion coefficients. In a more compact notation that implies summation over descriptor indices, the atomic energies can be written as
\begin{equation}
\label{eq18}
E^{\rm L}_i(\bm \eta) =  \sum_{p=1}^{N_{\rm e}} c^{(1)}_p D^{(1)}_{ip} +  \sum_{k=1}^{N_{\rm d}^{(2)}} c^{(2)}_k D^{(2)}_{ik} + \sum_{m=1}^{N_{\rm d}^{(3)}} c^{(3)}_m D^{(3)}_{im}   
\end{equation}
where $N_{\rm d}^{(2)} = N_{\rm b}^{(2)} N_{\rm e} (N_{\rm e}+1)/2$ and $N_{\rm d}^{(3)} = N_{\rm b}^{(3)} N_{\rm e}^2 (N_{\rm e}+1)/2$ are the number of two-body and three-body descriptors, respectively. The one-body descriptors are defined as follows
\begin{equation}
\label{eq2a}
D_{ip}^{(1)} =  \left\{
\begin{array}{ll}
1, & \mbox{if } Z_i = p \\
0, & \mbox{if } Z_i \neq p
\end{array} 
\right.   
\end{equation}
for $1 \le i \le N, 1 \le p \le N_{\rm e}$. The one-body descriptors are independent of atom positions, but dependent on atom types. 
}

{
In actual implementation, we calculate the potential energy and atomic forces without computing and storing the descriptors. We note from  (\ref{eq12c}), (\ref{eq15}), and (\ref{eq16}) that the atomic energies can be expressed in terms of the orthogonal basis functions as
\begin{multline}
E^{\rm L}_i(\bm \eta) =   c^{(1)}_{Z_i}  + \displaystyle  \sum_{j} \sum_{m=1}^{N_{\rm b}^{(2)}}  c^{(2)}_{m l(Z_i, Z_j)} U^{(2)}_m(r_{ij},  \bm \eta) \ + \\
\displaystyle \sum_{j} \sum_{k} \sum_{m=1}^{N_{\rm r}} \sum_{n=1}^{N_{\rm a}}  c^{(3)}_{m n \ell(Z_i, Z_j,Z_k)}  U^{(3)}_{mn}(r_{ij}, r_{ik}, \theta_{ijk}, \bm \eta), 
\end{multline}
where $l(Z_i, Z_j)$ and $\ell(Z_i, Z_j,Z_k)$ are the index mappings defined in (\ref{eq12d}) and (\ref{eq15b}), respectively.  In order to calculate the atomic forces, we need to compute the derivatives of the atomic energies $E_i^{\rm L}(\bm \eta)$ with respect to $r_{ij}$, $r_{ik}$, and $\theta_{ijk}$. This requires us to compute the derivatives of the orthogonal basis functions in (\ref{eq120q}) and (\ref{eq140}) with $O(NN_{\rm n}^2N_{\rm b}^{(3)})$ as discussed in the previous section. Hence, the overall computational cost of the linear POD potential is $O(N N_{\rm n}^2 N_{\rm b}^{(3)})$. The cost is linear in the number of atoms and the number of three-body basis functions, but quadratic in the number of neighbors. 
}

\revise{
The potential energy surface is then obtained by summing  atomic energies $E_i^{\rm L}$ for all atoms $i$ in the system as follows
\begin{equation}
\label{eq19dq}
 E^{\rm L} = \sum_{p=1}^{N_{\rm e}} c^{(1)}_p d^{(1)}_{p} + \sum_{k=1}^{N_{\rm d}^{(2)}} c^{(2)}_k d_k^{(2)} + \sum_{m=1}^{N_{\rm d}^{(3)}} c^{(3)}_m d_m^{(3)} ,
\end{equation}
where  
\begin{equation}
\label{eq19dw}
d_{p}^{(1)} = \sum_{i=1}^N D_{ip}^{(1)}, \quad d_{k}^{(2)} = \sum_{i=1}^N D_{ik}^{(2)}, \quad d_{m}^{(3)} = \sum_{i=1}^N D_{im}^{(3)},
\end{equation}
are one-body, two-body, and three-body global descriptors, respectively. The atomic forces are thus given by
\begin{equation}
\label{eq19eq}
 \bm F^{\rm L} = - \sum_{k=1}^{N_{\rm d}^{(2)}} c^{(2)}_k \nabla d_k^{(2)} - \sum_{m=1}^{N_{\rm d}^{(3)}} c^{(3)}_m \nabla d_m^{(3)} 
\end{equation}
where the symbol $\nabla$ denotes the derivatives of the global descriptors with respect to atom positions. 
}

\subsection{The quadratic POD potential}

%The quadratic descriptors  are four-body because they involve central atom $i$ together with three neighbors $j, k$ and $l$. quadratic per-atom descriptors

\revise{To construct a quadratic POD potential, we introduce a set of quadratic per-atom descriptors 
\begin{equation}
\label{eq19}
 D^{(2*3)}_{ikm}(\bm R, \bm Z, \bm \eta) = \frac{1}{2N}\left( D^{(2)}_{ik}  d_{m}^{(3)} + D^{(3)}_{im}   d_{k}^{(2)} \right)
\end{equation}
for $1 \le i \le N, 1 \le k \le N_{\rm d}^{(2)}, 1 \le m \le N_{\rm d}^{(3)}$, where $d_{k}^{(2)}$ and $d_{m}^{(3)}$ are two-body and three-body global descriptors defined in (\ref{eq19dw}). Because the quadratic per-atom descriptors are defined in terms of the products of the local descriptors and the global descriptors, they depend on positions and chemical identities of all atoms in the system. Therefore, the quadratic per-atom descriptors encapsulate the global environments of the atomic configuration. The per-atom energies of the quadratic POD potential are defined as
\begin{equation}
\label{eq18ww}
E^{\rm Q}_i(\bm R, \bm Z, \bm \eta) =  E^{\rm L}_i  +   \sum_{k=1}^{N_{\rm d}^{(2)}} \sum_{m=1}^{N_{\rm d}^{(3)}} c^{(2*3)}_{km}  D^{(2*3)}_{ikm}(\bm R, \bm Z, \bm \eta) 
\end{equation}
where $E^{\rm L}_i$ defined in (\ref{eq18}) are the per-atom energies of the linear POD potential and $c^{(2*3)}_{km}$ are coefficients associated with the quadratic per-atom descriptors. While  $E^{\rm L}_i$ depend on positions and chemical identities of atoms  within the cut-off distance $r_{\rm max}$ from the central atom $i$,  $E^{\rm Q}_i$ depend on positions and chemical identities of all atoms in the system due to the quadratic per-atom descriptors.

The PES of the quadratic POD potential is  obtained by summing  atomic energies $E_i^{\rm Q}$ for all atoms $i$ in the system  as follows
\begin{equation}
\label{eq19d}
 E^{\rm Q} = \sum_{p=1}^{N_{\rm e}} c^{(1)}_p d^{(1)}_{p} + \sum_{k=1}^{N_{\rm d}^{(2)}} c^{(2)}_k d_k^{(2)} + \sum_{m=1}^{N_{\rm d}^{(3)}} c^{(3)}_m d_m^{(3)} + \sum_{k=1}^{N_{\rm d}^{(2)}} \sum_{m=1}^{N_{\rm d}^{(3)}} c^{(2*3)}_{km} d^{(2*3)}_{km} ,
\end{equation}
where the quadratic global descriptors are calculated as 
\begin{equation}
\label{eq19b}
 d^{(2*3)}_{km} = \sum_{i=1}^N D^{(2*3)}_{ikm} = \frac{1}{N} \left( \sum_{i=1}^N D^{(2)}_{ik} \right) \left( \sum_{i=1}^N D^{(3)}_{im} \right) = \frac{1}{N} d^{(2)}_{k} d^{(3)}_m, 
\end{equation}
for $1 \le k \le N_{\rm d}^{(2)}, 1 \le m \le N_{\rm d}^{(3)}$.  The atomic forces are calculated as
\begin{equation}
\label{eq19e}
 \bm F^{\rm Q}  = - \sum_{k=1}^{N_{\rm d}^{(2)}} c^{(2)}_k \nabla d_k^{(2)} - \sum_{m=1}^{N_{\rm d}^{(3)}} c^{(3)}_m \nabla d_m^{(3)} - \sum_{k=1}^{N_{\rm d}^{(2)}} \sum_{m=1}^{N_{\rm d}^{(3)}} c^{(2*3)}_{km}  \nabla d^{(2*3)}_{km} .
\end{equation}
Hence, the quadratic POD potential is the linear POD potential augmented with the additional terms due to the quadratic global descriptors .
}

\revise{In order to efficiently evaluate  the quadratic POD potential, we rewrite its PES as 
\begin{equation}
\label{eq37}
 E^{\rm Q} =  \sum_{p=1}^{N_{\rm e}} c^{(1)}_p d^{(1)}_{p} + \sum_{k=1}^{N_{\rm d}^{(2)}} \left(c^{(2)}_k + 0.5 b_k^{(2)} \right) d_k^{(2)} +  \sum_{m=1}^{N_{\rm d}^{(3)}} \left( c^{(3)}_m + 0.5 b_m^{(3)} \right) d_m^{(3)}  ,
\end{equation}
and the atomic forces as
\begin{equation}
\label{eq19f}
 \bm F^{\rm Q} = - \sum_{k=1}^{N_{\rm d}^{(2)}}  \left(c^{(2)}_k + b^{(2)}_k \right) \nabla d_k^{(2)} - \sum_{m=1}^{N_{\rm d}^{(3)}} \left( c^{(3)}_m + b^{(3)}_m  \right) \nabla d_m^{(3)} ,
\end{equation}
where
\begin{equation}
\label{eq38}
\begin{split}
b_k^{(2)} & = \frac{1}{N} \sum_{m=1}^{N_{\rm d}^{(3)}} c^{(2*3)}_{km} d_m^{(3)}, \quad k = 1,\ldots, N_{\rm d}^{(2)}, \\
b_m^{(3)} & = \frac{1}{N} \sum_{k=1}^{N_{\rm d}^{(2)}} c^{(2*3)}_{km} d_k^{(2)}, \quad m = 1,\ldots, N_{\rm d}^{(3)} .
\end{split}
\end{equation}
To calculate $b^{(2)}_k$ and $b^{(3)}_m$, we must compute the two-body and three-body global descriptors 
\begin{equation}
\label{eq42c}
\begin{split}
d^{(2)}_{m l(p, q) }(\bm \eta)  & =  \sum_{\{i | Z_i = p\}} \sum_{\{j | Z_j = q\}} U^{(2)}_m(r_{ij},  \bm \eta) \\ 
d^{(3)}_{m n \ell(p, q, s)}(\bm \eta)  & =  \sum_{\{i | Z_i = p\}} \sum_{\{j | Z_j = q\}} \sum_{\{k | Z_k = s\}} U^{(3)}_{mn}(r_{ij}, r_{ik}, \theta_{ijk}, \bm \eta) ,
\end{split}
\end{equation}
with an operation count of $O(NN_{\rm n}^2N_{\rm b}^{(3)})$. The additional cost of calculating $b^{(2)}_k$ and $b^{(3)}_m$ in (\ref{eq38}) is $O(N_{\rm d}^{(2)} N_{\rm d}^{(3)})$. Therefore, the computational cost of the quadratic POD potential is $O(NN_{\rm n}^2N_{\rm b}^{(3)} + N_{\rm d}^{(2)} N_{\rm d}^{(3)})$. If $N_{\rm d}^{(2)} N_{\rm d}^{(3)}$ is less than $NN_{\rm n}^2N_{\rm b}^{(3)}$, the computational complexity of the quadratic POD potential reduces to $O(NN_{\rm n}^2N_{\rm b}^{(3)} )$, which is exactly the computational cost of the linear POD potential. This should be the case for most practical applications involving molecular dynamics simulations of large number of atoms.
}

\revise{The quadratic potential, with its comprehensive set of descriptors and incorporation of higher body orders, has the potential for more precise predictions than the linear potential. It achieves this without a significant increase in computational expense for many relevant systems, maintaining the same cost profile as its linear counterpart.  Furthermore, it is possible for the quadratic potential to capture long-range interactions because its per-atom energies depend on all atoms in the system. In other words, the changes in the position of one atom affect the forces upon all atoms in the system. However, the quadratic descriptors do not inherit the notion of orthogonality from the two-body and three-body descriptors, and are incomplete in the sense that they do not span the systematic set of  four-body descriptors. Nonetheless, this approach can be augmented for higher-order interactions through the integration of three-body descriptors with themselves, potentially expanding the predictive power of the model.}

\subsection{Connection with the quadratic SNAP potential}

\revise{The quadratic POD potential  bears similarities with the quadratic SNAP potential \cite{Wood2018}. The main idea of the quadratic SNAP is to make new per-atom descriptors from the original per-atom descriptors as follows  
\begin{equation}
\label{eq19q}
 D^{(2*3)}_{ikm} =  D^{(2)}_{ik}  D^{(3)}_{im}, \quad  k = 1, \ldots, N_{\rm d}^{(2)}, m = 1,\ldots, N_{\rm d}^{(3)} .
\end{equation}
The SNAP descriptors and our quadratic per-atom descriptors (\ref{eq19}) are distinct in their dependency on atomic arrangements. Specifically, SNAP descriptors are influenced by the local environment surrounding a central atom $i$ focusing on immediate atomic neighbors. In contrast, our quadratic per-atom descriptors consider interactions involving all atoms within the system and thus provide a more global description of atomic interactions. The SNAP  global descriptors are  calculated as
\begin{equation}
\label{eq19bq}
 d^{(2*3)}_{km} = \sum_{i=1}^N D^{(2*3)}_{ikm} =  \sum_{i=1}^N D^{(2)}_{ik}  D^{(3)}_{im} .
\end{equation}
We see that the quadratic global descriptors of the quadratic SNAP potential are not defined as the products of the original global descriptors, whereas those of the quadratic POD potential are as shown by (\ref{eq19b}). As a result, the quadratic SNAP potential cannot be written as a linear combination of the original global descriptors, whereas the quadratic POD potential can be as shown by (\ref{eq37}). The quadratic SNAP potential is slightly more expensive than the linear SNAP potential because evaluating the quadratic descriptors and their derivatives incurs an increase in computational cost and memory storage.}

\section{Potential Fitting}

\subsection{Weighted least-squares regression}

Let $J$ be the number of training configurations, with $N_j$ being the number of atoms in the $j$th configuration. Let $\{E^{*}_j\}_{j=1}^{J}$ and $\{\bm F^{*}_j\}_{j=1}^{J}$ be the DFT energies and forces for $J$ configurations. Next, we calculate the global descriptors and their derivatives for all training  configurations. Let $d_{jm}, 1 \le m \le M,$ be the global descriptors associated with the $j$th configuration, where $M$ is the number of global descriptors. We then form a matrix $\bm A \in \mathbb{R}^{J \times M}$ with entries $A_{jm} = d_{jm}/ N_j$ for $j=1,\ldots,J$ and $m=1,\ldots,M$. Moreover, we form a matrix $\bm B \in \mathbb{R}^{\mathcal{N} \times M}$ by stacking the derivatives of the global descriptors for all training configurations from top to bottom, where $\mathcal{N} = 3\sum_{j=1}^{J} N_j$. The coefficient vector $\bm c$ of the linear expansion (\ref{eq18}) is found by solving the following least-squares problem  
\begin{equation}
\label{eq45}
\min_{\bm c \in \mathbb{R}^{M}}    \beta \|\bm A(\bm \eta) \bm c - \bar{\bm E}^{*} \|^2 + \|\bm B(\bm \eta) \bm c + \bm F^{*} \|^2, 
\end{equation}
where $\beta$ is a scalar to weight the energy contribution. Here $\bar{\bm E}^{*} \in \mathbb{R}^{J}$ is a vector of with entries $\bar{E}^{*}_j = E^{*}_j/N_j$ and $\bm F^{*}$ is a vector of $\mathcal{N}$ entries obtained by stacking $\{\bm F^{*}_j\}_{j=1}^{J}$ from top to bottom. 

{The training procedure is the same for both the linear and quadratic POD potentials. However, the quadratic POD potential is  more expensive to train because the fitting of the quadratic POD potential still requires us to calculate and store the quadratic global descriptors and their gradient. Furthermore, the quadratic POD potential may require more training data in order to prevent over-fitting. In order to reduce the cost of fitting the second POD potential and avoid over-fitting, we can use subsets of two-body and three-body PODs for constructing the new descriptors.}

\subsection{Optimization of hyperparameters}

The hyperparameters contain inner and outer cut-off distances for two-body and three-body interactions. They can affect the prediction performance of the resulting potential. We observe through our numerical experiences that longer cut-off distances do not necessarily translate to better performance. Furthermore, setting them to some heuristic values may lead to an interatomic potential that can be less accurate than the potential obtained by optimizing the cut-off distances. Thus, we solve the following constrained optimization problem
\begin{equation}
\label{eq46}
\begin{split}
& \min_{\bm \eta \in \Omega^{\bm \eta}, \bm c \in \mathbb{R}^M }    w \|\bm A(\bm \eta) \bm c(\bm \eta) - \bar{\bm E}^{*} \|_{\rm MAE} + (1-w) \|\bm B(\bm \eta) \bm c(\bm \eta) - \bm F^{*} \|_{\rm MAE} \\
& \mbox{subject to }  (\beta^2 \bm A^T \bm A + \bm B^T \bm B) \bm c = (\beta^2 \bm A^T \bar{\bm E}^{*} + \bm B^T \bm F^{*})
\end{split}
\end{equation}
where $w \in [0,1]$ is a given scalar weight, and $\|\cdot \|_{\rm MAE}$ is defined as
\begin{equation}
\|\bm e\|_{\rm MAE} = \frac1n \sum |e_i| .
\end{equation}
Note that $\Omega^{\bm \eta}$ is a domain in which the hyperparameters reside. The equality constraint in (\ref{eq46}) is obtained from the optimality condition of the least-squares problem (\ref{eq45}). 

{The optimization problem (\ref{eq46}) is generally nonlinear and non-convex. Instead of directly solving the problem (\ref{eq46}), we solve a surrogate problem by approximating the loss function in (\ref{eq46}) using polynomial tensor-product interpolation. We select $Q$ grid points $\{\bm \eta_q \in \Omega^{\bm \eta}\}_{q=1}^Q$ and compute the loss function at those grid points. We then fit the loss values at the grid points to a tensor-product multivariate polynomial of certain degree to construct a surrogate model of the loss function. Finally, the gradient descent algorithm is used to minimize the surrogate model to obtain the hyperparameters. In order to be able to reach the global minimum, we start our gradient descent algorithm with a  number of initial guesses randomly  sampled in $\Omega^{\bm \eta}$.} 
\section{Numerical Experiments}

 In this section, we assess the performance of  POD potentials for indium phosphide and titanium dioxide. The linear POD potential in Section 3.1 and the quadratic POD potential in Section 3.2  shall be refereed to as POD-I and POD-II, respectively. In addition, we present results with the ACE potential \cite{DUSSON2022110946} and the explicit multi-element SNAP potential \cite{Cusentino2020}. 
 
\subsection{Indium phosphide}

In \cite{Cusentino2020}, Cusentino et. al introduced an explicit multi-element SNAP  potential and demonstrated it for indium phosphide (InP) dataset generated using the Vienna Ab Initio Simulation Package (VASP). The InP dataset contains not only chemically unique environment but also high-energy defects which are intended to study radiation damage effects where collision cascades of sufficiently high energy leave behind high formation energy point defects. In addition to defect configurations, the dataset includes configurations for uniform expansion and compression (Equation of State), random cell
shape modifications (Shear group), and uniaxially strained
(Strain group) unit cells for zincblende crystal structure. In total, the dataset has 1894 configurations with atom counts per configuration ranging from 8 to 216. {The InP data set is split into a training set (80\%) and test set (20\%) in order to assess the performance of POD potentials.} 

\begin{figure}[htbp]
	\centering	\includegraphics[width=0.65\textwidth]{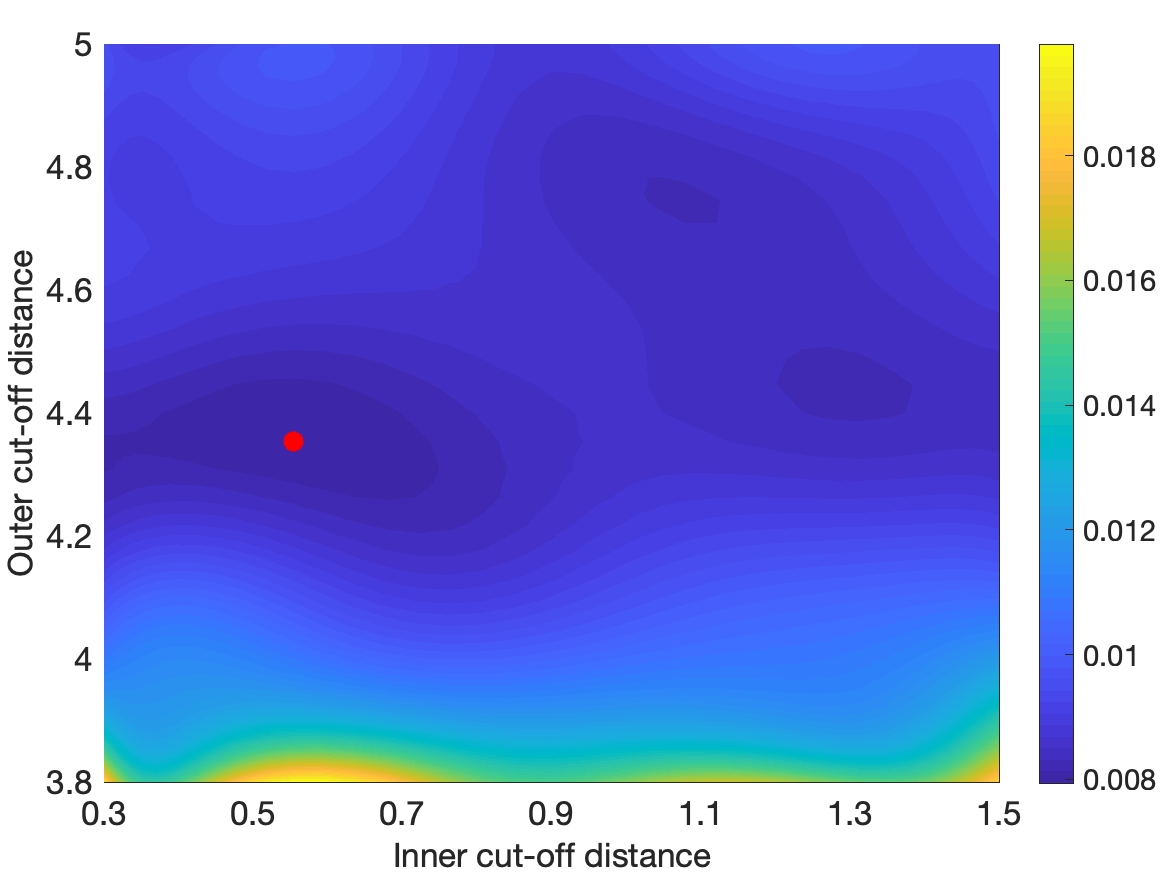}
	\caption{Surrogate loss function of the POD-I potential as a function of inner and outer cutoff distances for $N_{\rm d} = 170$. The red circle indicates the optimal value of the cutoff distances for which the surrogate loss function is minimum}
	\label{fig1}
\end{figure}

\begin{figure}[htbp]
\centering
\begin{subfigure}[b]{0.49\textwidth}
    \centering
    \includegraphics[width=\textwidth]{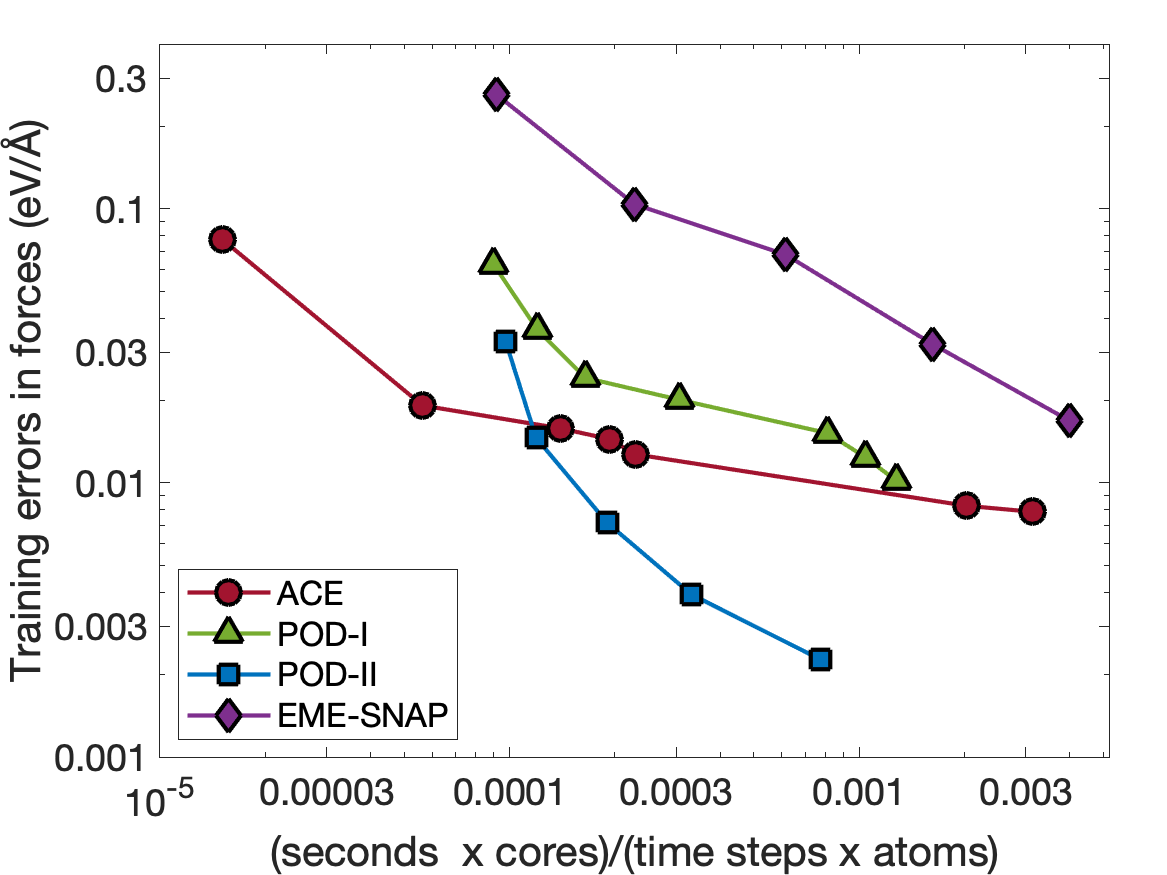}
\end{subfigure}
\hfill
\begin{subfigure}[b]{0.49\textwidth}
    \centering
    \includegraphics[width=\textwidth]{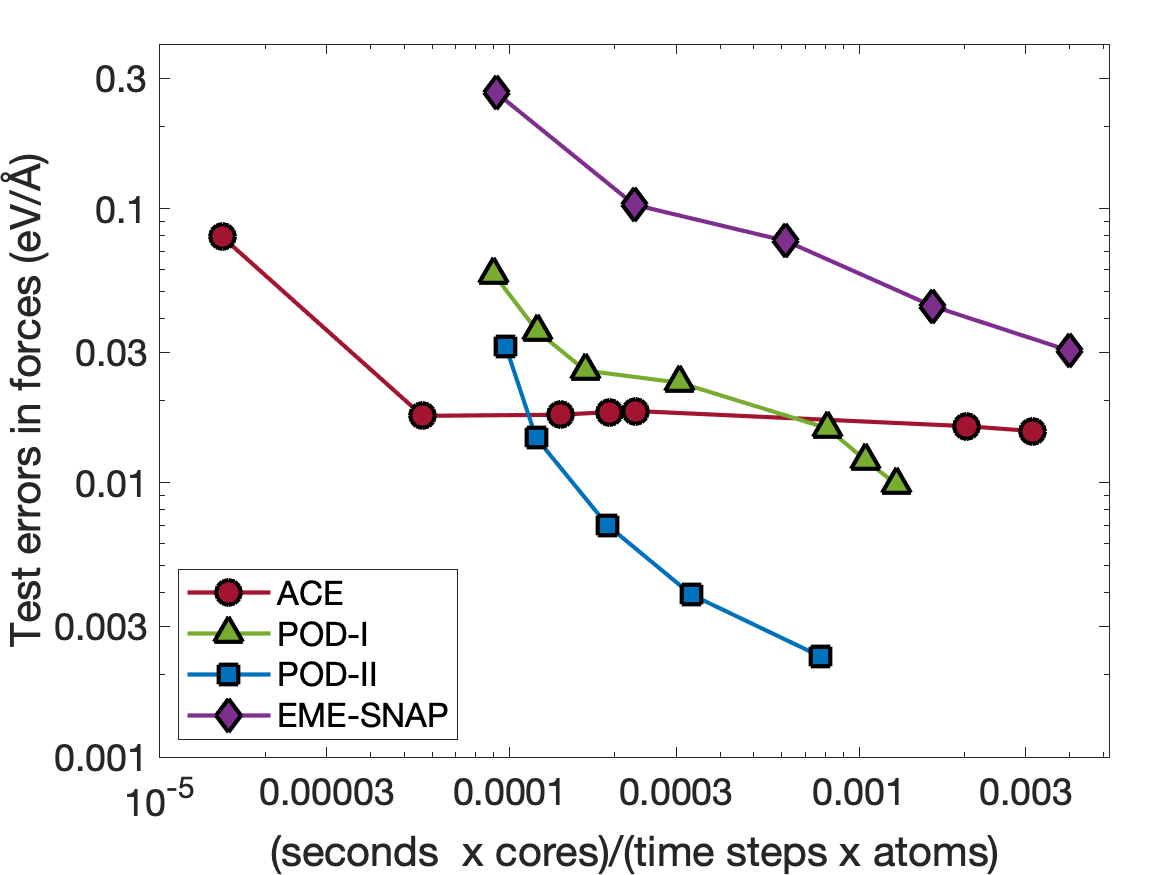}
\end{subfigure}
\caption{Force errors versus the computational cost of MD simulations for the InP system of 8000 atoms. MD simulations are performed using LAMMPS \cite{Thompson2022} on the 2.1 GHz Intel Broadwell E5-2695 v4 2S x 18C of 36 cores for ACE, EME-SNAP, POD-I, POD-II for the cases listed in Table \ref{tab1}.}
\label{fig42b} 
\end{figure}

{Table \ref{tab1} provides the  cutoff distances and the number of descriptors  for  POD-I, POD-II, and EME-SNAP, while Table \ref{tab1a} reports the polynomial degrees for ACE.\footnote{{POD-I, POD-II, SNAP, and ACE potentials are available at the Github repository https://github.com/cesmix-mit/pod-examples/tree/main/JCP2023\_InP to facilitate the reproduction of this work. The repository contains the input files for training these potentials and the output files for the MAE results reported in Tables \ref{tab1b} and \ref{tab1c}. Furthermore, there are input files for running MD simulations with the trained potentials and LAMMPS log files for the computational times reported in Figure \ref{fig42b}.}}  The number of descriptors for EME-SNAP is $N_{\rm d} = N_{\rm e}^4  (J+1)(J+2)(J+3/2)/3$. The number of descriptors for ACE depends on the body orders, the radial and angular degrees for each body order. For POD potentials, the optimization problem (\ref{eq46}) is solved with $\beta = 100$ and $w = 0.8$. Figure \ref{fig1} show the polynomial interpolation surrogate of the loss function for POD-I at $N_{\rm d} = 170$. The surrogate loss function is smooth with multiple local minima. We see that cutoff distances affect the value of the loss function and thus the accuracy of the resulting potential. Therefore, it is beneficial to optimize these hyperparameters. 
}
\begin{table}[htbp]
\centering
	\begin{tabular}{|c|ccc|ccc|ccc|}
		\cline{1-10}
		&
		 \multicolumn{3}{|c|}{\textbf{POD-I}} & 
		 \multicolumn{3}{c|}{\textbf{POD-II}} & 
		 \multicolumn{3}{c|}{\textbf{EME-SNAP}}\\
		\hline
		Case &  $N_{\rm d}$ & $r_{\min}$ & $r_{\max}$ & $N_{\rm d}$ & $r_{\min}$ & $r_{\max}$ & $N_{\rm d}$ & $J$ & $r_{\rm cut}$ \\
		\hline
		1 &   26 & 0.30 & 4.25 & 134 & 0.38 & 4.69 & 80 & 1 & 3.8 \\
		2 &   83 & 0.31 & 4.43 & 731 & 0.59 & 4.63 & 224 & 2 & 3.8\\
		3 &  170 & 0.56 & 4.37 & 2870 & 0.50 & 4.76  & 480 & 3 & 3.8\\
		4 &    314 & 0.36 & 4.80 & 7226 & 0.62 & 4.81 & 880 & 4 & 3.8\\
		5 &   515 & 0.65 & 4.93 & 16355 & 0.72 & 4.91 & 1456 & 5 & 3.8\\
  		6 &  764 & 0.65 & 4.93 & -- & -- & -- & 2240 & 6 & 3.8\\
            7 &  1058 & 0.65 & 4.93 & -- & -- & -- & 3264 & 7 & 3.8\\
		\hline
	\end{tabular}
	\caption{{The  cutoff distances  and number of descriptors for  EME-SNAP, POD-I, and POD-II. The EME-SNAP potential developed in \cite{Cusentino2020} is used here.}} 
	\label{tab1}
\end{table}

\begin{table}[htbp]
\centering
	\begin{tabular}{|c|ccccccccc|}	
		\hline
            Case & $N_d$ & $\ell_{\rm max}^1$ & $\ell_{\rm max}^2$ & $\ell_{\rm max}^3$ & $\ell_{\rm max}^4$ & $n_{\rm max}^1$ & $n_{\rm max}^2$ & $n_{\rm max}^3$ & $n_{\rm max}^4$  \\
            \hline
		1 & 70 & 1 & 1 &  1 & 0 & 2 & 2 & 1 & 0 \\
		2 & 252 & 1 & 2 &  3 & 0 & 8 & 3 & 1 & 0 \\
		3 & 446 & 1 & 2 &  4 & 0 & 16 & 4 & 1 & 0 \\
		4 &  738 & 1 & 3 &  5 & 0 & 12 & 5 & 1 & 0 \\
		5 & 1190 & 1 & 4 &  6 & 0 & 8 & 6 & 2 & 0 \\
  		6 & 3496 & 1 & 4 &  6 & 1 & 8 & 6 & 2 & 1 \\
            7 & 6046 & 1 & 4 &  6 & 2 & 8 & 6 & 2 & 2 \\
		\hline
	\end{tabular}
	\caption{{The number of descriptors, angular degrees, and radial degrees for ACE potentials. Here $\ell_{\rm max}^r$ and $n_{\rm max}^r$ refer to the maximum polynomial of degree of spherical harmonics and radial basis functions for the $(r+1)$-body descriptors, respectively. For all cases, the outer cut-off radii are $r_{\rm max}^{\rm InIn} = 5.790$,  $r_{\rm max}^{\rm InP} = 5.007$, $r_{\rm max}^{\rm PP} = 4.224$, while the inner cut-off radii are $r_{\rm min}^{\rm InIn} = 1.705$,  $r_{\rm min}^{\rm InP} = 1.403$, $r_{\rm min}^{\rm PP} = 1.100$. The ACE potentials are trained by using the FitSNAP package \cite{rohskopf2023fitsnap} and the ML-PACE package \cite{Lysogorskiy2021}.}} 
	\label{tab1a}
\end{table}

{Table \ref{tab1b} provides mean absolute errors (MAEs) in energies predicted using ACE, EME-SNAP, POD potentials, while Table \ref{tab1c} shows MAEs in forces. We see that the training errors decrease as the number of descriptors increases. Therefore, the training errors are correlated to the number of descriptors for all potentials. For POD-I and POD-II, the test errors are similar to the training errors. For ACE and EME-SNAP, the test errors in energies are considerably larger than the training errors in energies. In the first four cases, the test errors in forces are similar to the training errors in forces for ACE and EME-SNAP. In the last two cases,  the test errors in forces are about two times greater than the training errors in forces for ACE and EME-SNAP. The training errors for ACE are smaller than the training errors for POD-I, whereas the test errors for ACE are larger than the test errors for POD-I. We observe that POD-II has significantly lower errors than the other potentials.}

\begin{table}[htbp]
\small
\centering
	\begin{tabular}{|c|cc|cc|cc|cc|}
		\cline{1-9}
		\textbf{} &
		 \multicolumn{2}{|c|}{\textbf{ACE}} & \multicolumn{2}{c|}{\textbf{POD-I}} & 
		 \multicolumn{2}{c|}{\textbf{POD-II}} & 
		 \multicolumn{2}{c|}{\textbf{EME-SNAP}}\\
		\hline
		Case & training & test & training & test & training & test & training & test  \\
		\hline
1  &  27.05  &  99.72  &  21.18  &  18.78  &  4.08  &  3.90  &  64.75  &  160.15  \\  
 2  &  3.17  &  11.58  &  4.13  &  4.25  &  0.95  &  0.96  &  10.01  &  49.32  \\  
 3  &  1.66  &  12.02  &  2.88  &  2.82  &  0.38  &  0.40  &  3.16  &  29.64  \\  
 4  &  1.26  &  9.76  &  1.61  &  1.63  &  0.17  &  0.24  &  1.45  &  15.51  \\  
 5  &  0.90  &  7.73  &  1.01  &  1.02  &  0.07  &  0.40  &  0.47  &  9.46  \\  
 6  &  0.67  &  17.77  &  0.57  &  0.61  &  --  &  --  &  0.32  &  19.48  \\  
 7  &  0.65  &  8.29  &  0.34  &  0.39  &  --  &  --  &  0.22  &  47.07  \\   
		\hline
	\end{tabular}
	\caption{{Mean absolute errors (MAEs) in energies for ACE, EME-SNAP, POD-I, POD-II potentials. The unit of MAEs in energies is meV/atom.}} 
	\label{tab1b}
\end{table}

\begin{table}[htbp]
\small
\centering
	\begin{tabular}{|c|cc|cc|cc|cc|}
		\cline{1-9}
		\textbf{} &
		 \multicolumn{2}{|c|}{\textbf{ACE}} & \multicolumn{2}{c|}{\textbf{POD-I}} & 
		 \multicolumn{2}{c|}{\textbf{POD-II}} & 
		 \multicolumn{2}{c|}{\textbf{EME-SNAP}}\\
		\hline
		Case & training & test & training & test & training & test & training & test  \\
		\hline
1  &  77.51  &  79.19  &  62.65  &  57.88  &  32.70  &  31.48  &  261.77  &  265.46  \\  
 2  &  19.18  &  17.58  &  36.20  &  35.73  &  14.63  &  14.63  &  104.15  &  103.62  \\  
 3  &  15.82  &  17.74  &  24.18  &  25.72  &  7.15  &  7.01  &  68.03  &  76.44  \\  
 4  &  14.47  &  18.15  &  20.11  &  23.22  &  3.92  &  3.91  &  32.04  &  44.15  \\  
 5  &  12.72  &  18.31  &  15.21  &  15.83  &  2.27  &  2.33  &  16.89  &  30.39  \\  
 6  &  8.25  &  16.11  &  12.41  &  12.13  &  --  &  --  &  12.74  &  25.42  \\  
 7  &  7.85  &  15.47  &  10.19  &  9.89  &  --  &  --  &  7.90  &  21.44  \\  
		\hline
	\end{tabular}
	\caption{{Mean absolute errors (MAEs) in forces for ACE, EME-SNAP, POD-I, POD-II potentials. The unit of MAEs in forces is meV/\AA.}} 
	\label{tab1c}
\end{table}

{Figure \ref{fig42b}
shows the trade-off between computational cost and force errors for cases listed in Table \ref{tab1}. The computational cost is measured in terms of number of cores multiplied by the run time in seconds  per time steps per atoms for molecular dynamics (MD) simulations. These MD simulations are performed using LAMMPS \cite{Thompson2022} on the 2.1 GHz Intel Broadwell E5-2695 v4 2S x 18C processor of 36 cores with $10 \times 10 \times 10$ bulk supercell containing 8000 InP atoms. We see that  POD-II has the same computational cost as POD-I, while having considerably lower force errors than POD-I. This is because POD-II has a greater number of descriptors than POD-I for exactly the same computational complexity. Meanwhile, ACE, POD-I, and POD-II are many times faster than EME-SNAP for the same accuracy. For the level of accuracy of force errors less than 0.015 eV/\AA, POD-II  outperforms the other potentials in the sense that POD-II has the lowest computational cost for such level of accuracy. However, ACE performs better than POD-II for the accuracy of force errors of more than 0.015 eV/\AA.}

\revise{Due to the quadratic per-atom descriptors, the quadratic POD potential demonstrates significantly greater accuracy compared to the linear POD, SNAP, and ACE potentials. This enhanced accuracy largely stems from the global scope of the quadratic per-atom descriptors, which take into account the entire system rather than just local interactions. This global approach allows for a more detailed and accurate modeling of complex atomic behaviors and interactions. In contrast, because the linear POD, SNAP, and ACE potentials are short-range potentials, they may not be able to capture important global effects. This limitation highlights the advantage of the quadratic POD potential in environments where long-range interactions play a critical role in the behavior of the system.}

\begin{figure}[htbp]
	\centering
	\begin{subfigure}[b]{0.49\textwidth}
		\centering
		\includegraphics[width=\textwidth]{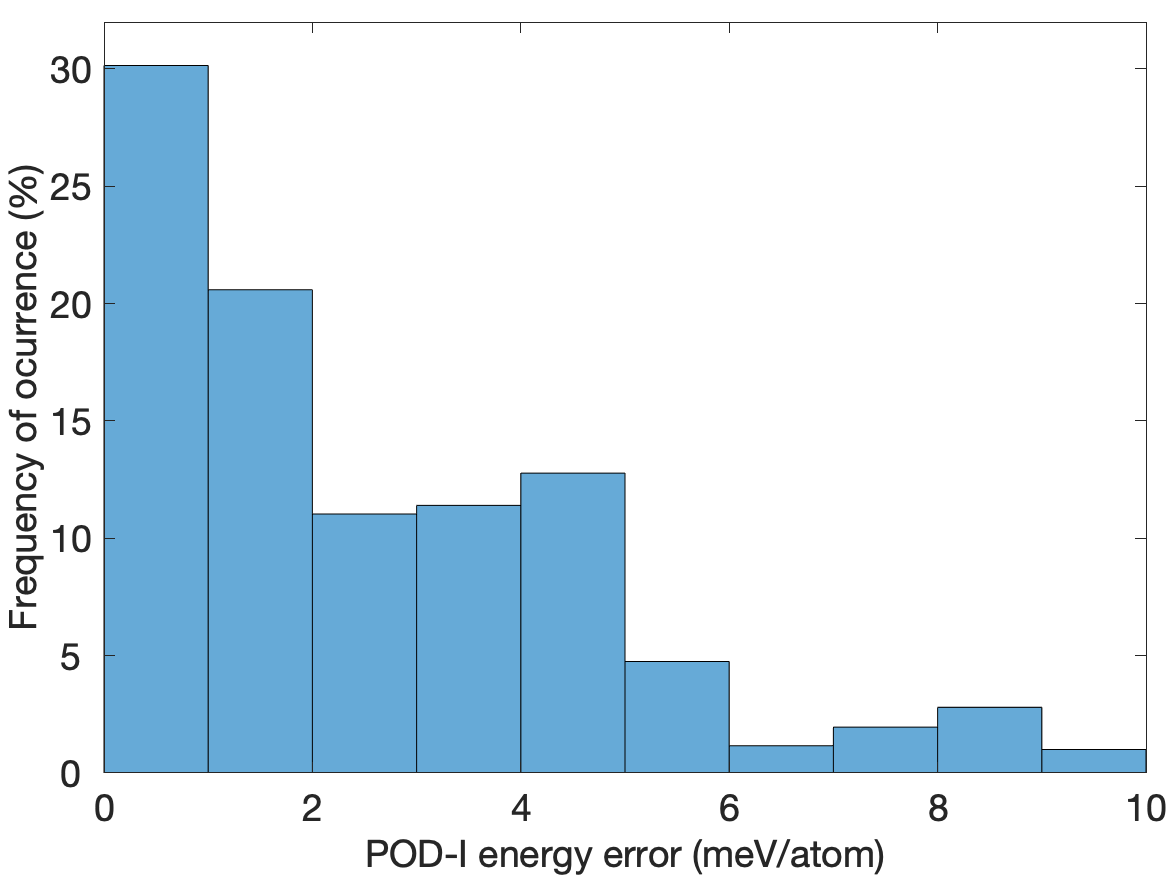}
	\end{subfigure}
	\hfill
	\begin{subfigure}[b]{0.49\textwidth}
		\centering
		\includegraphics[width=\textwidth]{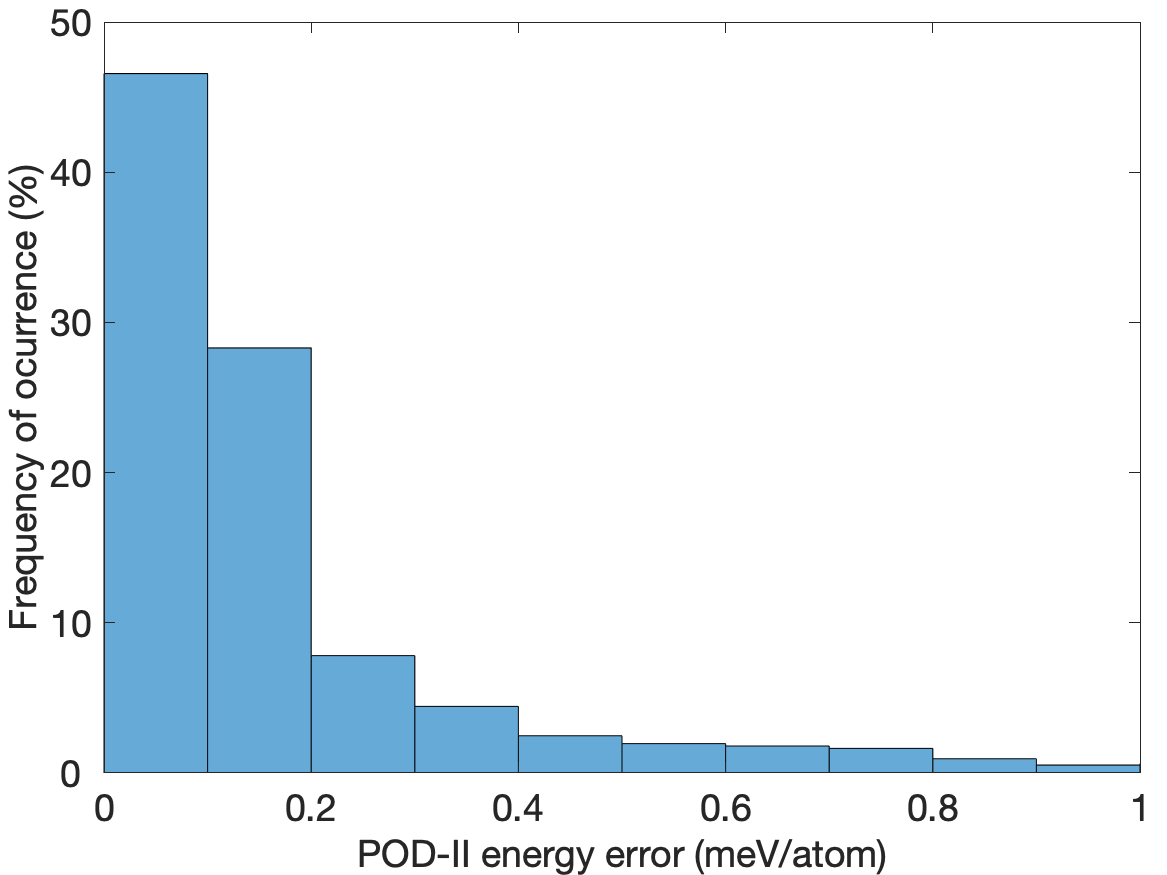}
	\end{subfigure}
	\caption{Histogram distribution of energy errors for all structures in the training set as predicted by POD-I (left) and POD-II (right) for Case 3.}
	\label{fig2bb}
\end{figure}

\begin{figure}[htbp]
	\centering
	\begin{subfigure}[b]{0.49\textwidth}
		\centering
		\includegraphics[width=\textwidth]{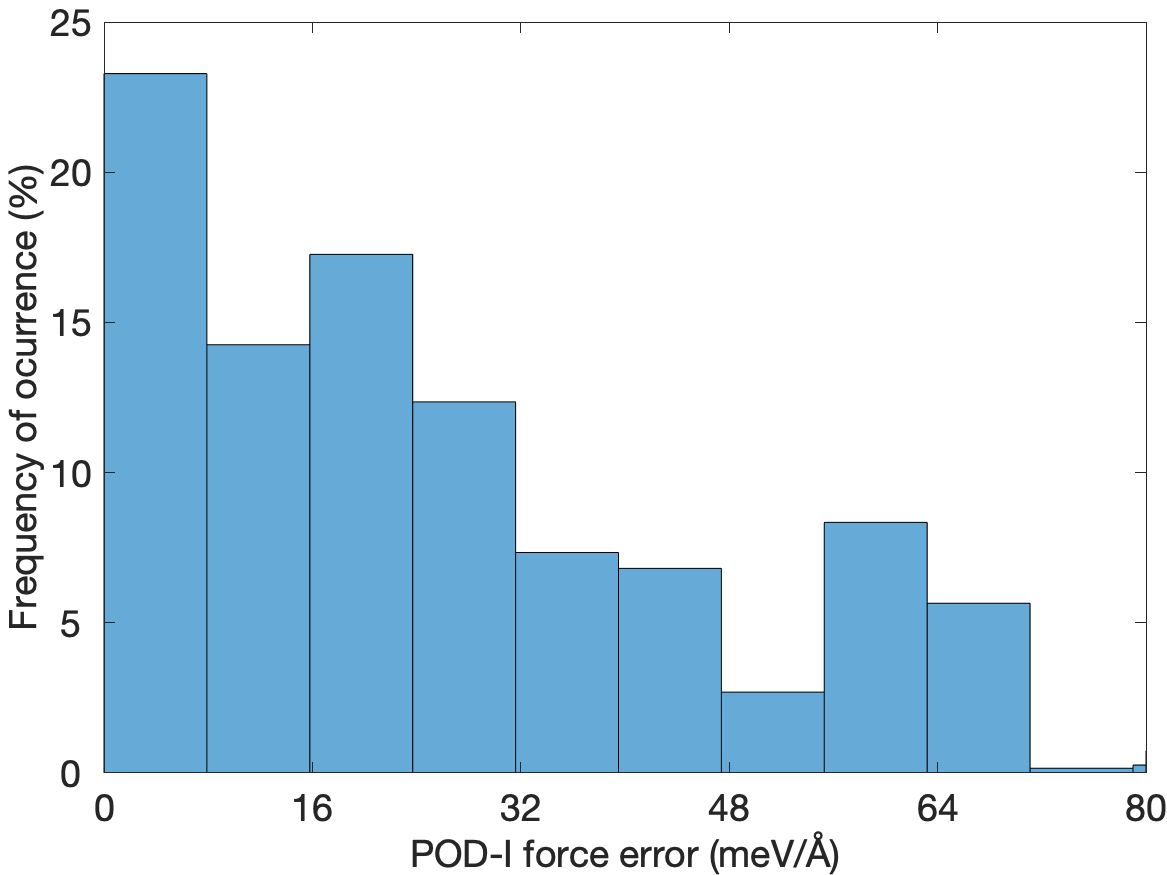}
	\end{subfigure}
	\hfill
	\begin{subfigure}[b]{0.49\textwidth}
		\centering
		\includegraphics[width=\textwidth]{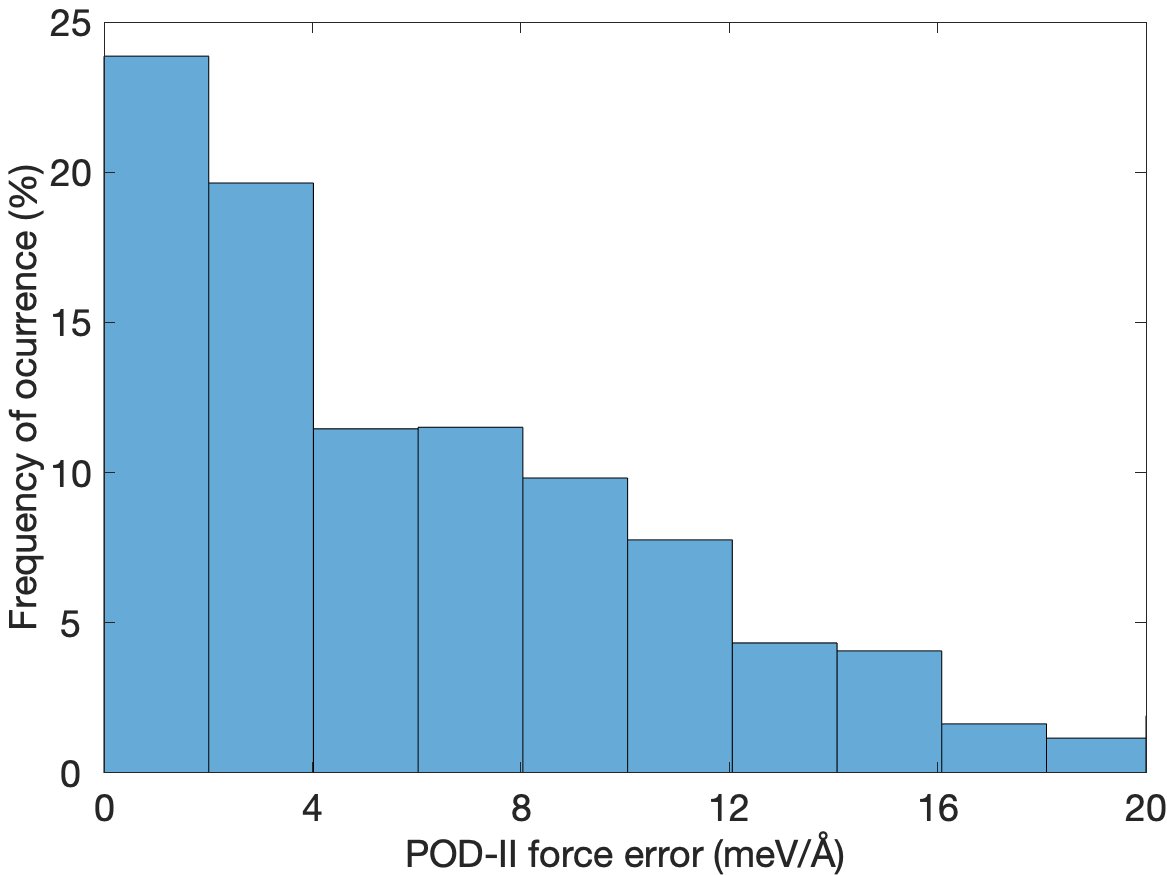}
	\end{subfigure}
	\caption{Histogram distribution of force errors for all structures in the training set  as predicted by POD-I (left) and POD-II (right) for Case 3.}
	\label{fig2bc}
\end{figure}

Figure \ref{fig2bb} displays the histogram of mean absolute energy errors for all the training structures as predicted by POD-I and POD-II  for Case 3. For POD-I, more than 50\% of the structures has an energy error of less than 2 meV/atom, while 99\% of the structures are predicted with an energy error of less than 10 meV/atom. For POD-II, almost all the structures are predicted with an energy error of less than 1 meV/atom, while 75\% of the structures has an energy error of less than 0.2 meV/atom. Although these error statistics for energy prediction are very good, for practical purposes it is crucial that atomic forces are predicted accurately. No matter how accurate the energy prediction is, the potential is unsuitable for molecular dynamics simulations if the predicted forces are inaccurate. As shown in Figure \ref{fig2bc},  more than 95\% of the structures are predicted with a force error of less than 80 meV/\AA \ for POD-I and 20 meV/\AA \ for POD-II. At such level of accuracy in force prediction, POD-I is acceptable for molecular dynamics simulations, while POD-II is far superior to POD-I.
 
Point defects are created when atoms become vacant at lattice sites (vacancy defect),  occupy locations in the crystal structure at which there are usually not an atom (interstitial defect), or exchange positions with other atoms of different types (antisite defect). {Table \ref{tab2d} (respectively,  Table \ref{tab2e}) shows the errors in energies (respectively, forces) for the following point defects created from an equilibrium configuration of 64 atoms: $\mbox{In}_{\rm a} \mbox{P}_{\rm a}$, $\mbox{In}_{\rm v} \mbox{P}_{\rm v}$, $\mbox{In}_{\rm a}$, $\mbox{P}_{\rm a}$, $\mbox{In}_{\rm i}$, $\mbox{P}_{\rm i}$, $\mbox{In}_{\rm v}$, and $\mbox{P}_{\rm v}$, where subscripts correspond to vacancy(v), interstital(i) and anti-site(a)  defects. All the potentials predict these point defects very accurately.  EME-SNAP predicts the vacancy defects more accurately than the other point defects. POD-II has the lowest errors for almost  point defects.} 

\begin{table}[htbp]
\centering
\small
	\begin{tabular}{|cc|cc|cc|cc|cc|cc|}
		\cline{1-10}
		 %\textbf{Group} &
   \multicolumn{2}{|c}{\textbf{Defect Type}} &	 
	 \multicolumn{2}{|c|}{\textbf{ACE}} & 
		 \multicolumn{2}{c|}{\textbf{POD-I}} &
		 \multicolumn{2}{c|}{\textbf{POD-II}} & \multicolumn{2}{c|}{\textbf{EME-SNAP}} \\
   Name  & $N_{\rm{atom}}$ & train & test & train & test & train & test & train & test \\
		\cline{1-10}
$\mbox{In}_{\rm a} \mbox{P}_{\rm a}$ & 64 & 0.32  &  0.43  &  6.84  &  6.89  &  0.03  &  0.02  &  0.02  &  0.06  \\  
$\mbox{In}_{\rm v} \mbox{P}_{\rm v}$ & 62  & 4.78  &  3.86  &  2.15  &  2.52  &  0.02  &  0.01  &  0.04  &  0.05  \\  
$\mbox{In}_{\rm a}$ & 64 & 0.61  &  0.92  &  1.45  &  1.41  &  0.12  &  0.13  &  0.15  &  0.25  \\  
$\mbox{P}_{\rm a}$ & 64 & 1.31  &  0.89  &  3.30  &  3.36  &  0.08  &  0.13  &  0.59  &  8.69  \\  
$\mbox{In}_{\rm i}$ & 65 &  0.61  &  0.70  &  0.66  &  0.59  &  0.02  &  0.02  &  0.06  &  0.22  \\  
$\mbox{P}_{\rm i}$ & 65  & 0.97  &  2.97  &  0.53  &  0.50  &  0.02  &  0.02  &  0.24  &  1.00  \\  
$\mbox{In}_{\rm v}$ & 63 & 0.08  &  0.17  &  3.80  &  3.72  &  0.03  &  0.02  &  0.03  &  0.02  \\  
$\mbox{P}_{\rm v}$ & 63  & 3.40  &  5.33  &  0.27  &  0.24  &  0.03  &  0.04  &  0.03  &  0.02  \\  
		\hline
	\end{tabular}
	\caption{{MAEs in energies (meV/atom)  for different point defects for ACE, EME-SNAP, POD-I, POD-II potentials with Case 5. The point defects are created from an equilibrium configuration of 64 atoms by inserting atoms  (interstitial defects), removing atoms (vacancy defects), or exchanging atoms of different types (antisite defects). Subscripts correspond to vacancy(v), interstitial(i) and antisite(a) point defects.}} 
	\label{tab2d}
\end{table}

\begin{table}[htbp]
\centering
\small
	\begin{tabular}{|cc|cc|cc|cc|cc|cc|}
		\cline{1-10}
		 %\textbf{Group} &
   \multicolumn{2}{|c}{\textbf{Defect Type}} &	 
	 \multicolumn{2}{|c|}{\textbf{ACE}} & 
		 \multicolumn{2}{c|}{\textbf{POD-I}} &
		 \multicolumn{2}{c|}{\textbf{POD-II}} & \multicolumn{2}{c|}{\textbf{EME-SNAP}} \\
   Name  & $N_{\rm{atom}}$ & train & test & train & test & train & test & train & test \\
		\cline{1-10}
$\mbox{In}_{\rm a} \mbox{P}_{\rm a}$ & 64 & 9.93  &  9.49  &  24.07  &  23.87  &  2.47  &  2.01  &  13.24  &  12.14  \\   
$\mbox{In}_{\rm v} \mbox{P}_{\rm v}$ & 62  & 25.03  &  24.48  &  34.36  &  40.90  &  2.97  &  2.99  &  7.15  &  13.86  \\  
$\mbox{In}_{\rm a}$ & 64 & 14.29  &  14.13  &  9.93  &  10.72  &  1.58  &  1.64  &  32.46  &  32.82  \\  
$\mbox{P}_{\rm a}$ & 64 & 15.54  &  21.57  &  26.42  &  29.69  &  2.30  &  2.49  &  24.51  &  85.14  \\  
$\mbox{In}_{\rm i}$ & 65 &  13.24  &  38.37  &  28.46  &  28.47  &  2.44  &  2.37  &  16.20  &  68.10  \\   
$\mbox{P}_{\rm i}$ & 65  & 19.74  &  32.95  &  21.84  &  21.56  &  2.21  &  2.20  &  16.18  &  39.11  \\  
$\mbox{In}_{\rm v}$ & 63 &8.50  &  7.82  &  17.41  &  15.85  &  1.78  &  2.00  &  0.45  &  0.93  \\  
$\mbox{P}_{\rm v}$ & 63  & 11.76  &  11.19  &  18.08  &  14.56  &  1.67  &  1.44  &  2.38  &  2.36  \\    
		\hline
	\end{tabular}
	\caption{{MAEs in forces (meV/\AA)  for different point defects for ACE, EME-SNAP, POD-I, POD-II potentials with Case 5.}} 
	\label{tab2e}
\end{table}

\begin{figure}[htbp]
\centering
	\begin{subfigure}[b]{0.32\textwidth}
		\centering
		\includegraphics[width=\textwidth]{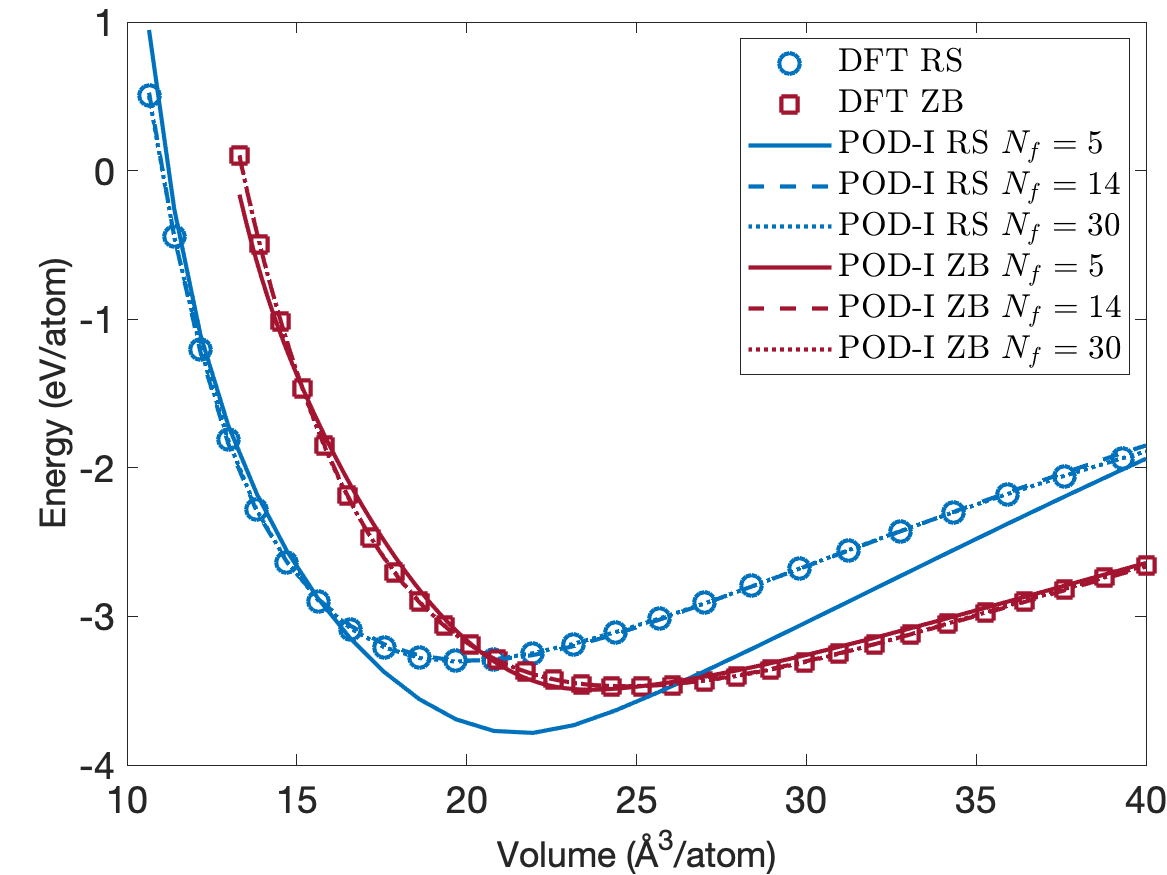}
		\caption{POD-I}
	\end{subfigure}
	\hfill
	\begin{subfigure}[b]{0.32\textwidth}
		\centering
		\includegraphics[width=\textwidth]{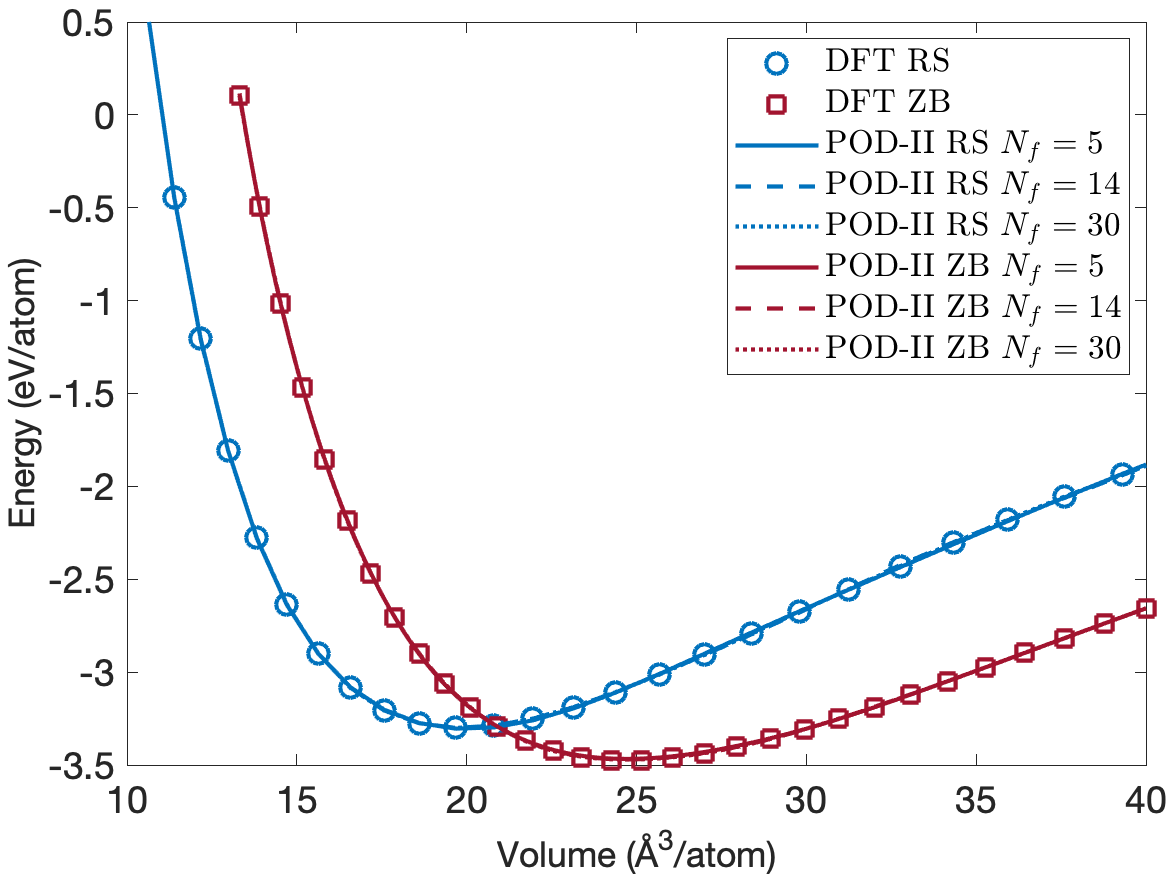}
		\caption{POD-II}
	\end{subfigure}
	\begin{subfigure}[b]{0.32\textwidth}
		\centering
		\includegraphics[width=\textwidth]{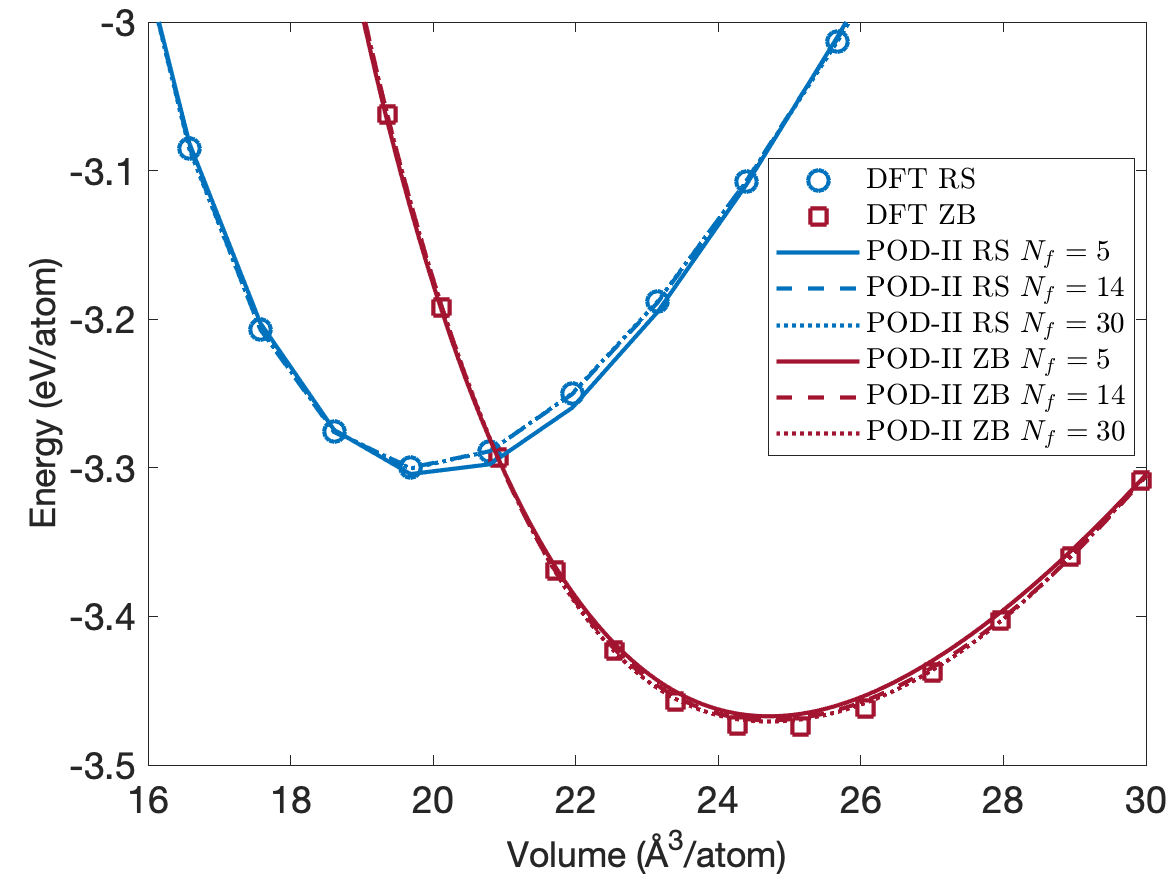}
		\caption{Close-up view }
	\end{subfigure}
\caption{Energy per atom versus volume per atom for rocksalt (RS) and zincblende (ZB) crystal structures of the InP compound for POD-I and POD-II potentials. The right figure shows the close-up view of the curves near the cohesive energies. Here $N_{\rm f} = N_{\rm b}^{(2)} + N_{\rm b}^{(3)}$ is the total number of basis functions for POD potentials.}
\label{Ta:energy}	
\end{figure}

In addition to defect formation energies, we also study cohesive energies for different low-energy crystal structures. Figure \ref{Ta:energy} plots the energy per atom computed with DFT, POD-I, POD-II as a function of volume per atom for the rocksalt (RS) and zincblende (ZB) crystal structures. We see that the predicted cohesive energies are very close to the DFT cohesive energies for both the rocksalt (RS) and zincblende (ZB) crystal structures. Furthermore, POD-I and POD-II correctly predict ZB as the most stable structure and reproduce the experimental cohesive energy of -3.48 eV/atom at a volume of 24.4 $\mbox{\AA}^3$/atom \cite{Nichols1980}. The predicted cohesive energies for the RS structure match exactly the DFT value of -3.30eV/atom at a volume of 19.7 $\mbox{\AA}^3$/atom. While not plotted in Figure \ref{Ta:energy}, the predicted cohesive energies for the wurtzite ground state structure agree well with the DFT value of -3.45eV/atom at a volume of 25.1 $\mbox{\AA}^3$/atom. The potentials accurately predict the cohesive energies for the low-energy crystal structures.

In this study, the point defects of the InP compound were
the primary focus, since accurately reproducing the defect formation energies is essential for realistic simulations of radiation damage \cite{Cusentino2020}. We found that the quadratic POD potential significantly improves the accuracy of energy and force predictions when it is compared to the linear POD potential as well as ACE and EME-SNAP potentials. Specifically, it has approximately 10 smaller energy errors and 4 times smaller force errors than the linear POD potential. We also found that all the potentials are able to accurately predict the formation energies of many different defects  far beyond the chemical accuracy of DFT. Nevertheless, additional training data will be needed to improve these ML potentials for other target applications. For instance, adding liquid phase training data is useful to study melting of InP. 

\subsection{Titanium dioxide}

Titanium dioxide (TiO$_2$) is one of the most chemically stable, environmentally compatible, and functionally versatile metal oxides, thereby making itself an ideal compound to demonstrate interatomic potentials. Over the years, several empirical potentials have been developed for TiO$_2$, including rigid-ion models \cite{Matsui1991,Oliver1997}, polarizable models \cite{Swamy2000,Swamy2001},  ReaxFF potential \cite{Kim2013}, and COMB potential \cite{Cheng2014}. More recently, a number of neural network potentials have been constructed for TiO$_2$, including artificial neural network (ANN) potential \cite{Artrith2016}, committee neural network (CNN) potential \cite{Schran2021}, and deep neural network (DNN) potential \cite{Andrade2020}.  Herein we would like to demonstrate the performance of ACE, SNAP, and POD potentials for TiO$_2$ on publicly available DFT data set \cite{Artrith2016}. 

\begin{figure}[htbp]
	\centering
	\includegraphics[width=0.85\textwidth]{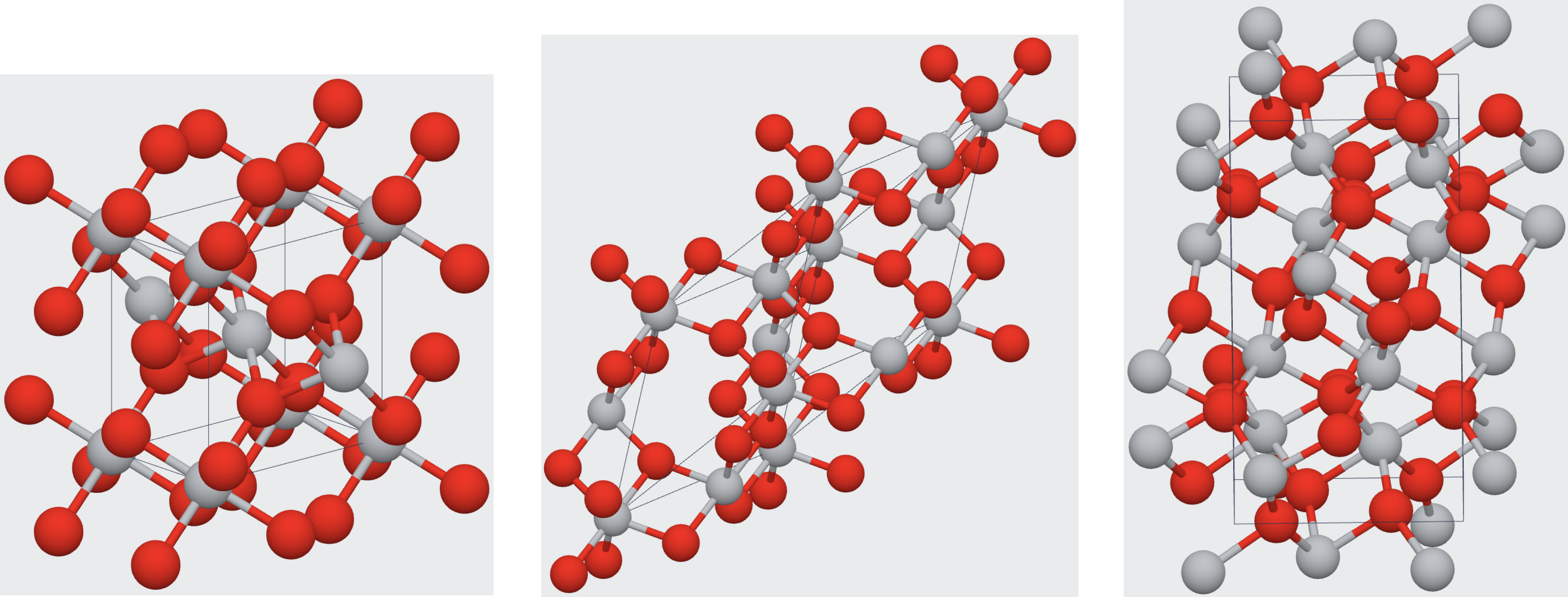}
	\caption{Unit cells of the three TiO$_2$ polymorphs: rutile (left),  anatase (middle), and (c) brookite (right). Red spheres indicate oxygen atoms and grey spheres are titanium atoms.}
	\label{figTiO2}
\end{figure}

Natural TiO$_2$ minerals occur in three polymorphs: rutile (tetragonal), anatase. (tetragonal), and brookite (orthorhombic). The unit cells of these polymorphs are shown in Figure \ref{figTiO2}. In \cite{Artrith2016}, a diverse DFT data set was generated to train an ANN potential for TiO$_2$ using an iterative refinement method. An initial set of reference structures was generated by distorting ideal rutile, anatase, and brookite structures around the ground state. In addition, supercell structures with oxygen vacancies were also included in the initial data set. Subsequently, additional reference structures were generated by performing short molecular dynamics simulations at various temperatures. This process was repeated until the newly generated structures were  found to be accurately represented by the ANN potential. The final data set comprised a total of 7815 structures containing between 6 and 95 atoms.  We refer to \cite{Artrith2016} for additional information about the DFT data set for TiO$_2$.

\begin{figure}[htbp]
	\centering
	\begin{subfigure}[ht]{0.49\textwidth}
		\centering
		\includegraphics[height=4.5cm]{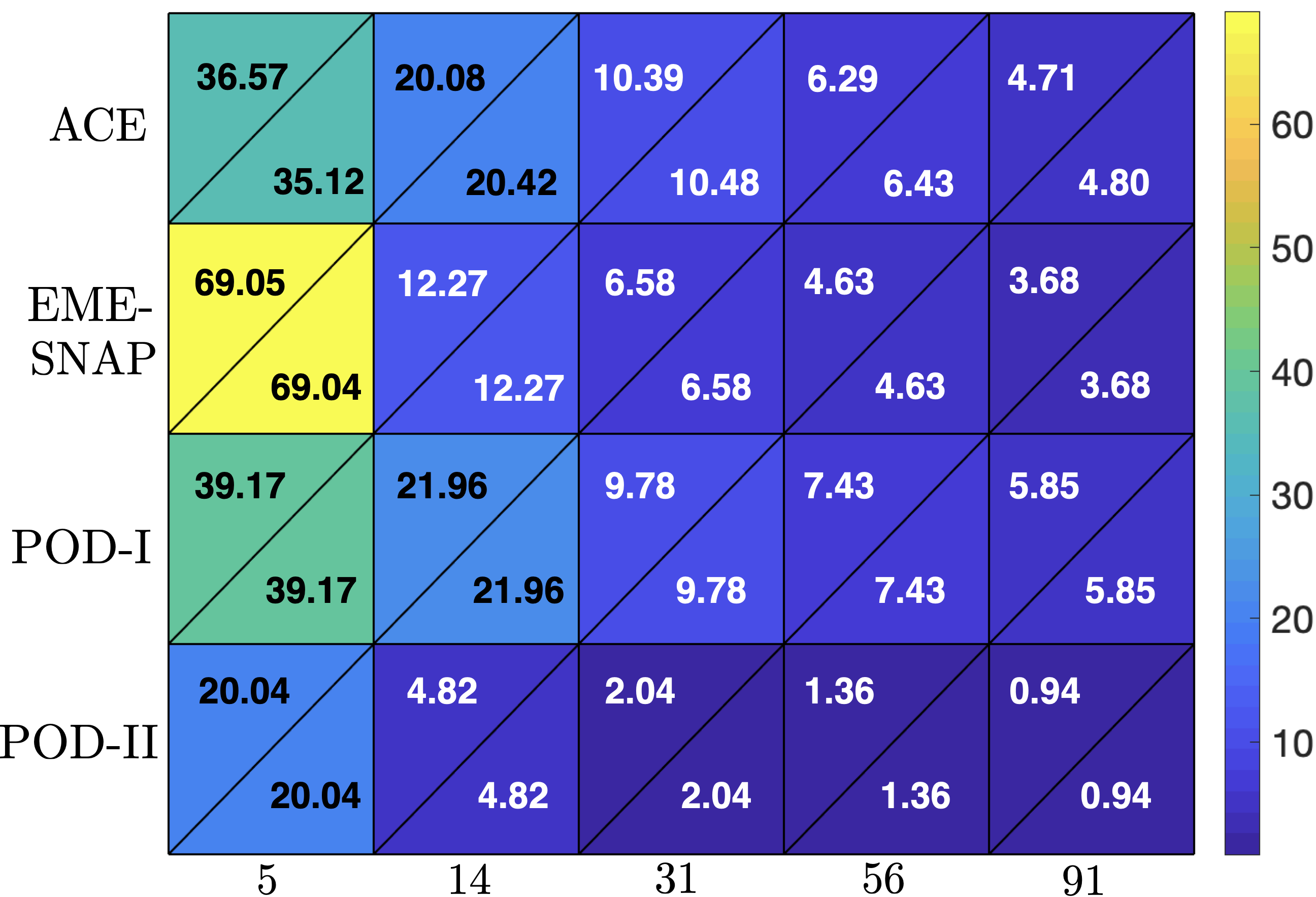}
		\caption{Energy errors (meV/atom)}
	\end{subfigure}
	\hfill
	\begin{subfigure}[ht]{0.49\textwidth}
		\centering
		\includegraphics[height=4.5cm]{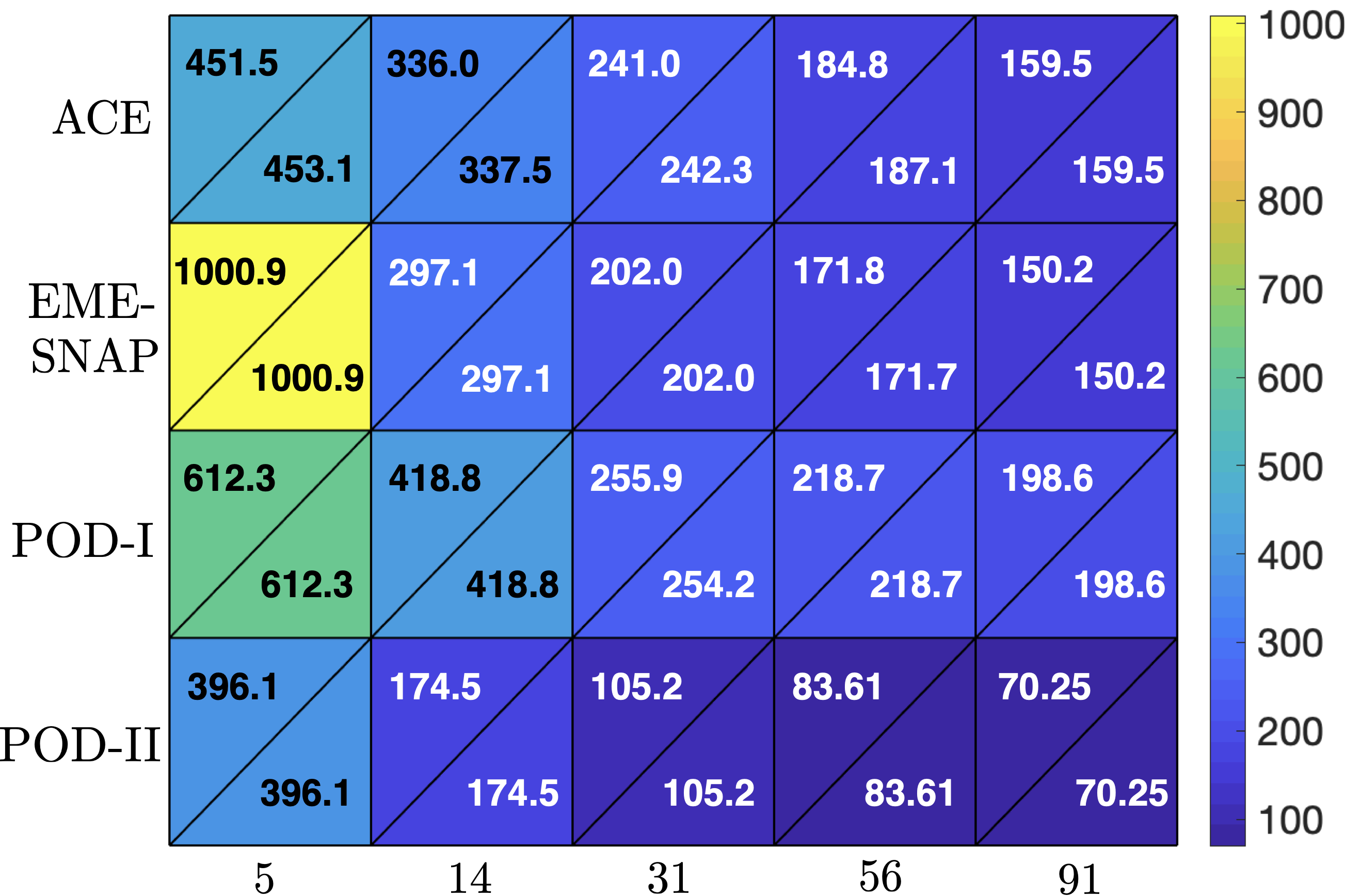}
		\caption{Force errors (meV/\AA)}
	\end{subfigure}
	\caption{Mean absolute errors in energies (a) and forces (b) for $N_{\rm f} = 5, 14, 31, 56,  91$ for TiO$_2$. The upper left and lower right triangles within each cell represent training and test errors, respectively. For POD potentials, the number of basis functions $N_{\rm f}$ is defined as the sum of the two-body and three-body basis functions. For ACE and EME-SNAP, the number of basis functions $N_{\rm f}$ is defined as the number of bispectrum components per atom for one-element systems.}
	\label{fig2TiO2}
\end{figure}
 
We use 50\% of the data set (3908 training structures) to fit ACE, SNAP, and POD potentials, while employing the rest of the data set (3907 test structures) as an independent testing set to assess the accuracy of the potentials and detect overfitting. A common way to detect overfitting is to compare the accuracy of the model for test structures (test errors) to the accuracy of the model for training structures (training errors). The training and test errors are shown in Figure \ref{fig2TiO2}. The  test errors are almost identical to the training errors for ACE, SNAP, and POD-I, POD-II potentials. The convergence of both energy and force errors is quite fast when $N_{\rm f}$ is less than 30. {The errors decrease quite slowly as $N_{\rm f}$ increases above 30.} In terms of accuracy comparison, ACE is slightly more accurate than POD-I, while SNAP is more accurate than ACE.  POD-II is the most accurate potential since it has the lowest energy and force errors. Indeed, POD-II  with $N_{\rm f} = 30$ is even more accurate than POD-I, ACE and SNAP with $N_{\rm f} = 91$. For $N_{\rm f} = 91$, POD-II predicts the mean absolute errors of about 0.94 meV/atom and 70 meV/\AA \ which are several times smaller than those predicted by the other potentials.

\revise{We emphasize that while we compare our approach to the linear ACE and SNAP potentials, it is conceivable that a nonlinear ACE potential \cite{Lysogorskiy2021} or an equivariant message passing neural network potential like MACE \cite{Batatia2022} may offer comparable improvements. Further exploration into these potentials could provide additional insights into the effectiveness of different computational approaches for modeling complex interactions within systems.}

\begin{figure}[htbp]
	\centering
	\begin{subfigure}[ht]{0.49\textwidth}
		\centering
		\includegraphics[width=0.95\textwidth]{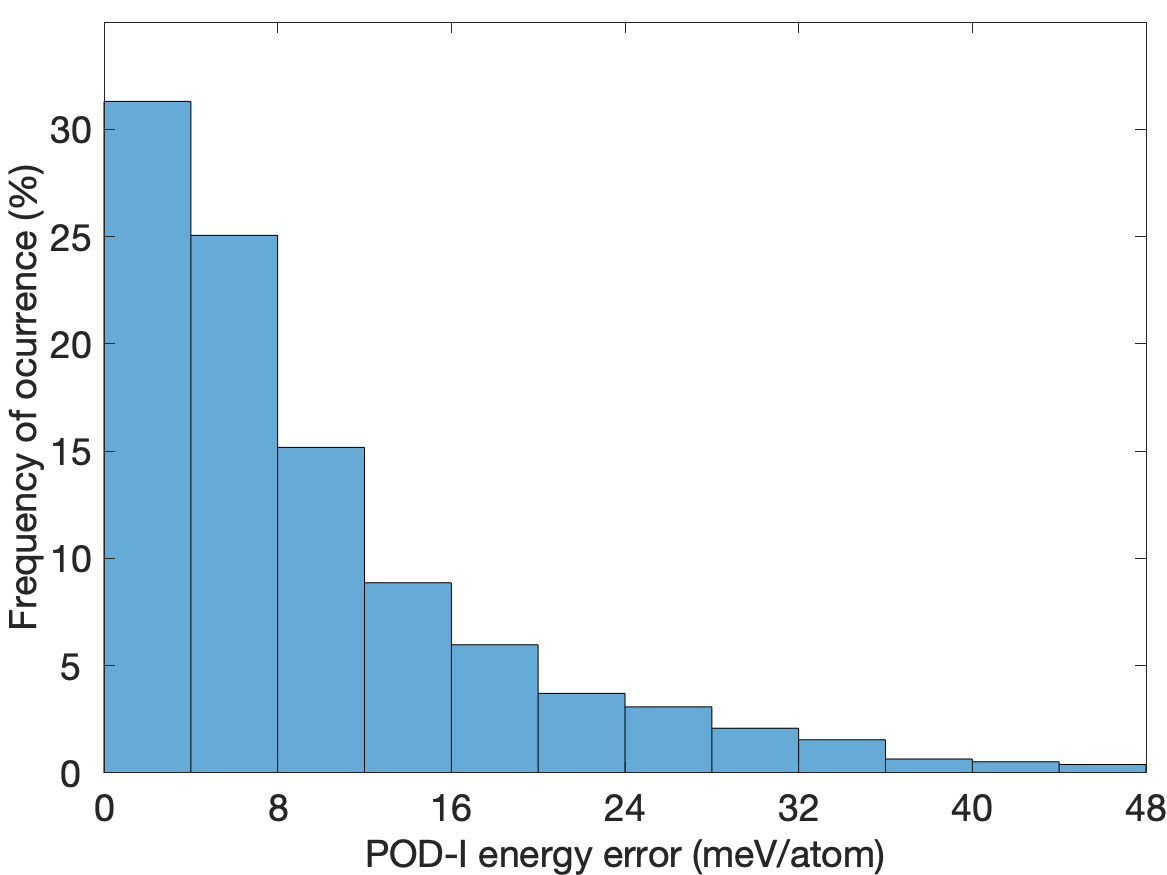}
	\end{subfigure}
	\hfill
	\begin{subfigure}[ht]{0.49\textwidth}
		\centering
		\includegraphics[width=0.95\textwidth]{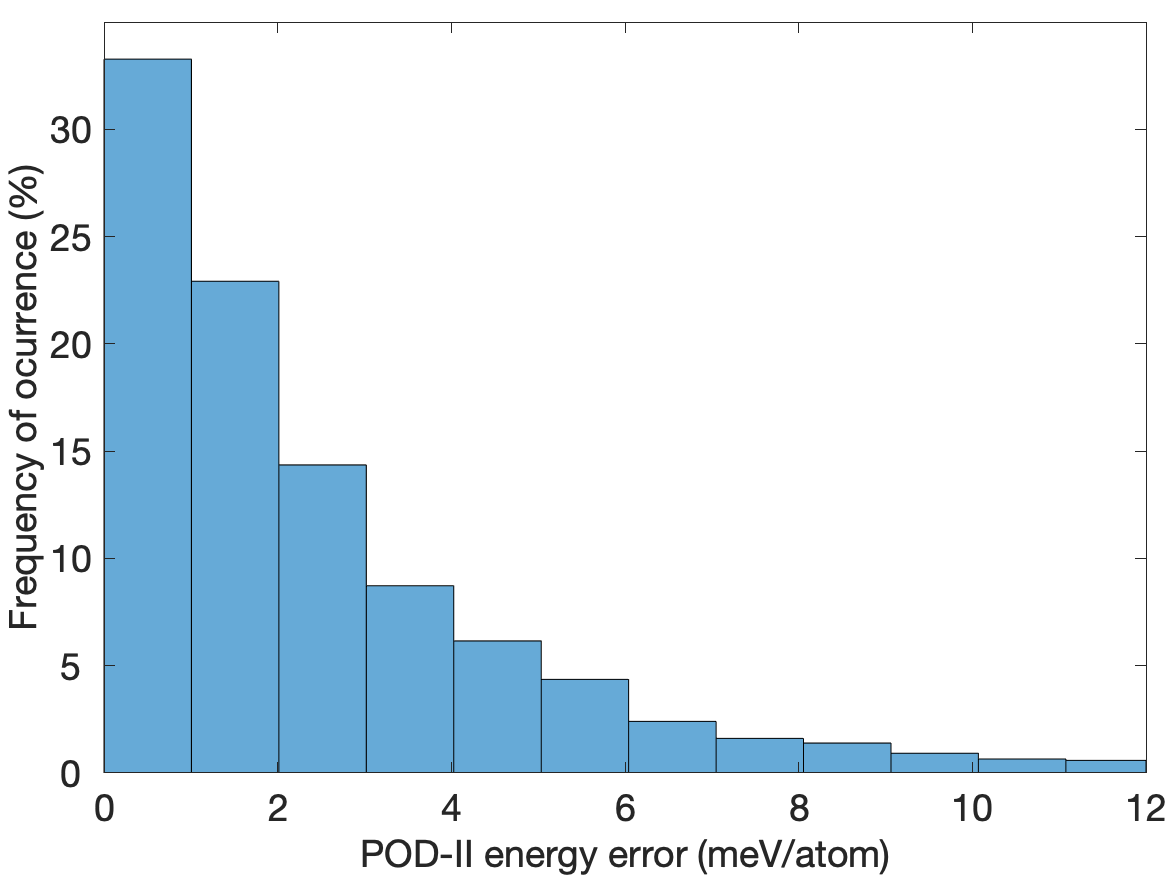}
	\end{subfigure}
	\caption{Histogram distribution of energy errors for all structures in the test set as predicted by POD-I (left) and POD-II (right) using $N_{\rm f} = 30$ basis functions.}
	\label{fig3TiO2}
\end{figure}

Figure \ref{fig3TiO2} displays the histogram of mean absolute energy errors for all the test structures as predicted by POD-I and POD-II using $N_{\rm f} = 30$ basis functions. For POD-I, more than 55\% of the structures has an energy error of less than 8 meV/atom, while 99\% of the structures are predicted with an energy error of less than 50 meV/atom. For POD-II, almost all the structures are predicted with an energy error of less than 12 meV/atom, while 55\% of the structures has an energy error of less than 2 meV/atom. The errors predicted by POD-II are several times less than those by POD-I. Hence, POD-II provides significantly more accurate prediction than POD-I while having the same computational cost. 

 \begin{figure}[htbp]
	\centering
	\begin{subfigure}[ht]{0.49\textwidth}
		\centering
		\includegraphics[width=0.9\textwidth]{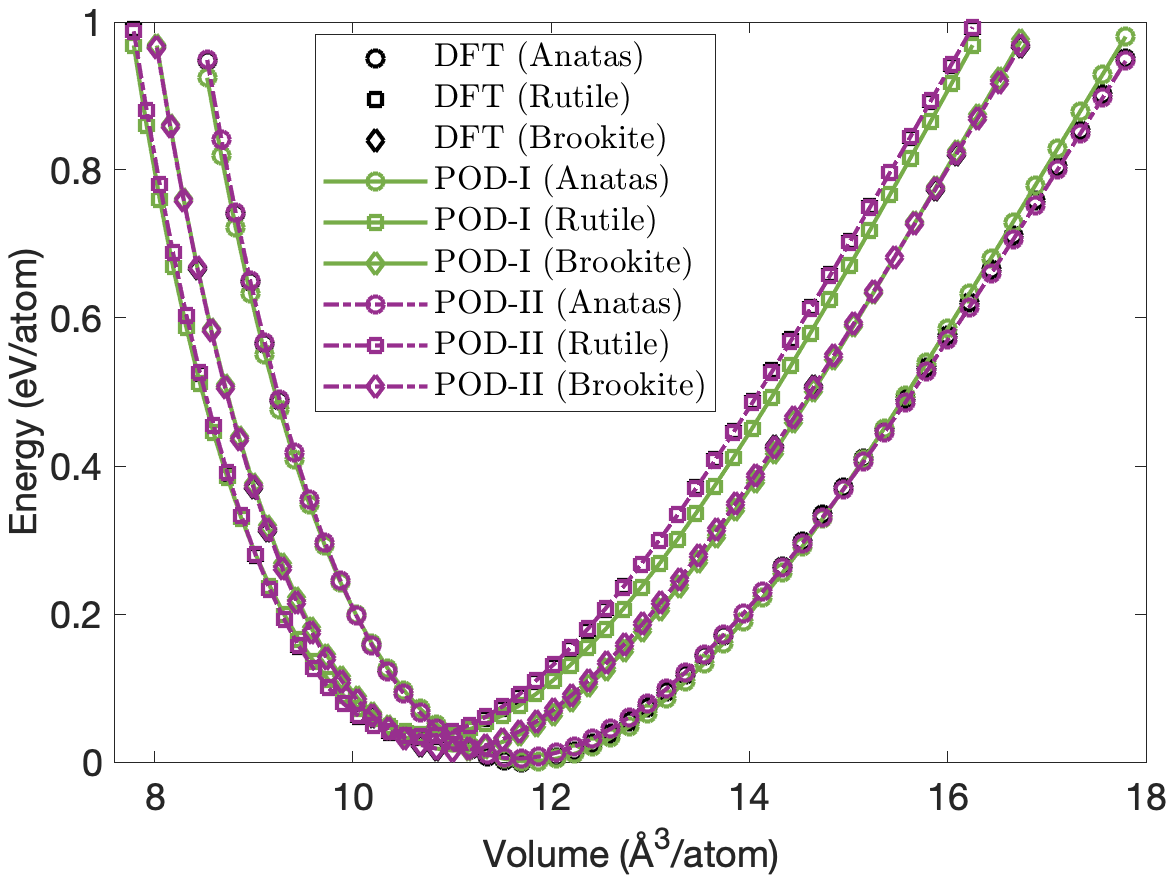}
		\caption{Energy versus volume}
	\end{subfigure}
	\hfill
	\begin{subfigure}[ht]{0.49\textwidth}
		\centering
		\includegraphics[width=0.9\textwidth]{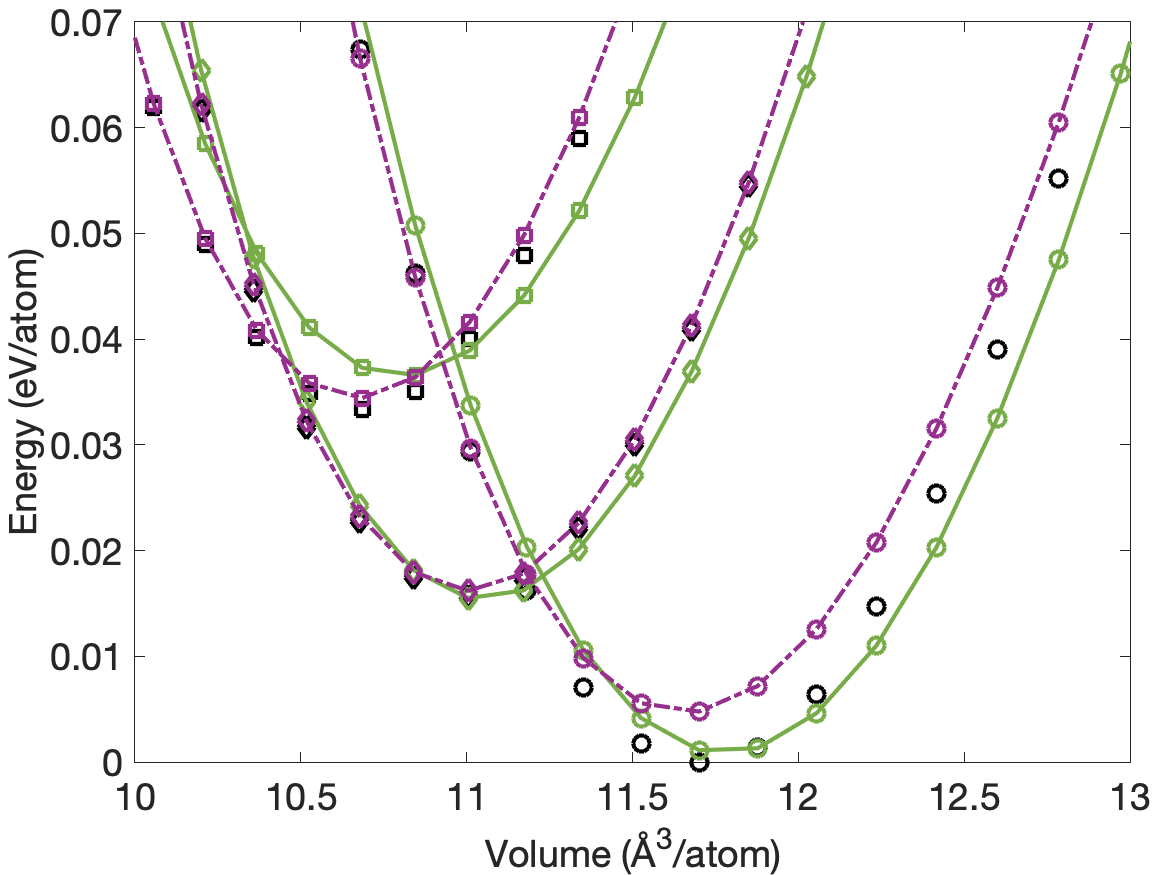}
		\caption{Close-up view}
	\end{subfigure}
	\caption{Energy per atom versus volume per atom for anatase, rutile, and brooktile crystal structures of the TiO$_2$ compound for DFT, and POD-I, POD-II potentials with $N_{\rm f} = 30$. The right figure shows the close-up view of the curves near the cohesive energies.}
	\label{fig4TiO2}
\end{figure}
 
 Figure \ref{fig4TiO2} plots the energy per atom computed with DFT, POD-I, and POD-II as a function of volume per atom for the anatase, rutile, and brooktile crystal structures. Both POD-I and POD-II potentials smoothly reproduce the energy as function of the lattice volume of rutile, anatase, and brookite. It is crucial that the energies of similar structures are reproduced smoothly. If the energy errors on the order of 2 meV/atom were completely stochastic, the  potential energy surface would be rugged and thus unsuitable for MD simulations. 
 Table \ref{tab2TiO2} summarizes the cohesive  energies and unit cell volumes for the rutile, anatase, and brookite structures. The agreement between the values predicted by the POD potentials and their DFT references
is excellent.

 \begin{table}[htbp]
\centering
\small
	\begin{tabular}{|c|cc|cc|cc|}
		\cline{1-7}
		Crystal  & \multicolumn{2}{|c|}{\textbf{DFT}} & \multicolumn{2}{c|}{\textbf{POD-I}} & 
		 \multicolumn{2}{c|}{\textbf{POD-II}}\\
		\cline{2-7}
		phase & $E_0$ & $V_0$ & $E_0$ & $V_0$ & $E_0$ & $V_0$ \\
		\hline
Anatase & 0.000 & 11.71 & 0.001 & 10.70 & 0.005 & 10.70 \\ 
 Rutile & 0.033 & 10.68 & 0.037 & 10.85 & 0.034 & 10.68 \\ 
 Brooktile & 0.016 & 11.01 & 0.016 & 11.01 & 0.016 & 11.01 \\ 
		\hline
	\end{tabular}
	\caption{Cohesive energies per atom $E_0$ (eV/atom)  and unit cell volumes per atom $V_0$ (\AA/atom) for three different TiO$_2$ phases as computed using density-functional theory (DFT) and POD potentials with $N_{\rm f} = 30$.}
	\label{tab2TiO2}
\end{table}

In this study, the phase energies of the TiO$_2$ compound were the primary focus, since smoothly reproducing the phase energies is essential for MD simulations. The quadratic POD potential provided more accurate predictions than the other potentials. The reference data set was restricted to periodic bulk structures. As a consequence, the constructed TiO$_2$ potentials can therefore not be expected to describe surface models (slab structures) or liquid phase accurately. Additional training data will be needed to further improve these  potentials for other target applications.

\section{Conclusions}

{We have developed proper orthogonal descriptors for multi-element chemical systems based on the orthogonal expansion of the radial and angular components of the parametrized potentials. In our work, the parametrized potentials are designed to  provide a rich and diverse representation of two-body and three-body interactions.  We compose the proper orthogonal descriptors in two different ways to construct two different interatomic potentials. The first POD potential expresses the energy of each atom as a linear combination of proper orthogonal descriptors, while the second POD potential expresses the energy as a linear and quadratic combination of the descriptors.

We have demonstrated the two POD potentials for indium phosphide and titanium dioxide, and compared their performance to that of multi-element SNAP and ACE potentials. For all potentials, increasing the number of basis functions usually improves prediction of energies and forces. The quadratic POD potential was shown to provide much more accurate prediction of energies and forces than the linear POD potential at the same computational cost. The improvement stems from the  composition of two-body and three-body descriptors to construct four-body descriptors without increasing the computational complexity of the resulting potential. 

The ideas presented in this paper can be extended to both empirical and machine learning potentials. Recently, a more efficient form of the proper orthogonal descriptors is developed by employing the atom density formulation \cite{Nguyen2023b}. We would like to extend the work \cite{Nguyen2023b} to multi-element chemical systems by utilizing the presented ideas. The two-body orthogonal basis functions developed herein can be used as radial basis functions in power spectrum, bispectrum, SOAP and ACE descriptors. Instead of fitting the parameters of empirical potentials, one can use the present method to construct an orthogonal basis for the parametric manifold of empirical potentials. This approach may improve  the accuracy and transferability of existing empirical potentials.
}

\section*{Acknowledgements} \label{}

{I am grateful to Dr. Andrew Rohskopf at Sandia National Laboratories (SNL) for fitting  ACE potentials and running MD simulations. I would like to thank all members of the CESMIX center at MIT for fruitful and invaluable discussions leading me to the ideas presented in this work.  I would like to thank Dr. Axel Kohlmeyer at Temple University and Dr. Aidan Thompson (SNL) for fruitful discussions about LAMMPS implementation of POD potentials.
I would also like to thank the reviewers for their constructive comments. I  gratefully acknowledge the United States  Department of Energy under contract DE-NA0003965 and the Air Force Office of Scientific Research under Grant No. FA9550-22-1-0356 for supporting this work.}

 \bibliographystyle{elsarticle-num} 
 %\bibliography{cas-refs}
\bibliography{library}

\end{document}